\documentclass[amsmath,amssymb]{revtex4}

\usepackage{graphicx}
\usepackage{amscd,amsmath,amsthm,amsfonts,amssymb}
\def\[{\begin{equation}}
\def\]{\end{equation}}

\begin{document}
\title{Rogue wave patterns in the nonlinear Schr\"{o}dinger equation}
\author{Bo Yang and Jianke Yang}
\affiliation{Department of Mathematics and Statistics, University of Vermont, Burlington, VT 05405, USA}

\begin{abstract}
Rogue wave patterns in the nonlinear Schr\"{o}dinger equation are analytically studied. It is shown that when an internal parameter in the rogue waves (which controls the shape of initial weak perturbations to the uniform background) is large, these waves would exhibit clear geometric structures, which are formed by Peregrine waves in shapes such as triangle, pentagon, heptagon and nonagon, with a possible lower-order rogue wave at its center. These rogue patterns are analytically determined by the root structures of the Yablonskii-Vorob'ev polynomial hierarchy, and their orientations are controlled by the phase of the large parameter. It is also shown that when multiple internal parameters in the rogue waves are large but satisfy certain constraints, similar rogue patterns would still hold. Comparison between true rogue patterns and our analytical predictions shows excellent agreement.
\end{abstract}

\maketitle

\section{Introduction}
The name of rogue waves first appeared in oceanography, where it referred to large spontaneous and unexpected water wave excitations that are a threat even to big ships \cite{Ocean_rogue_review,Pelinovsky_book}. Later, their counterparts in optics were also reported \cite{Solli_Nature,Wabnitz_book}. Due to their physical importance, rogue waves have received intensive theoretical and experimental studies in the past decade. On the theoretical front, analytical expressions of rogue waves have been derived in a wide variety of integrable physical models, such as the nonlinear Schr\"{o}dinger (NLS) equation for wave-packet propagation in the ocean and optical systems \cite{Benney,Peregrine,AAS2009,DGKM2010,KAAN2011,GLML2012,OhtaJY2012}, the derivative NLS equations for circularly polarized nonlinear Alfv\'en waves in plasmas and short-pulse propagation in a frequency-doubling crystal
\cite{Kaup_Newell,KN_Alfven1,Wise2007,KN_rogue_2011,KN_rogue_2013,YangDNLS2019}, the Manakov equations for light transmission in randomly birefringent fibers \cite{Menyuk,BDCW2012,ManakovDark,LingGuoZhaoCNLS2014,Chen_Shihua2015,ZhaoGuoLingCNLS2016}, and the three-wave resonant interaction equations \cite{Ablowitz_book,BaroDegas2013,DegasLomba2013,ChenSCrespo2015,WangXChenY2015,ZhangYanWen2018}.
On the experimental front, various rogue wave solutions in the NLS equation and defocusing Manakov equations have been observed in water tanks and optical fibers \cite{Tank1,Tank2,Fiber1,Fiber2,Fiber3}. In these experiments, intimate knowledge of theoretical rogue wave solutions in the underlying nonlinear wave equations was utilized, which highlights the importance of theoretical developments on rogue waves for practical rogue wave verifications and predictions.

The study of rogue wave patterns is important as this information allows for the prediction of later rogue wave events from earlier wave forms. Although graphs of low-order rogue waves have been plotted for many integrable equations, and simple patterns such as triangles and rings have been reported, richer patterns arising from high-order rogue wave solutions have received little attention. For the NLS equation, preliminary investigations on rogue wave patterns were made in \cite{KAAN2011,HeFokas,KAAN2013} through Darboux transformation and numerical simulations. It was observed in \cite{KAAN2011} that if a $N$-th order rogue wave exhibits a single-shell ring structure, then the center of the ring is a $(N-2)$-th order rogue wave. This observation was explained analytically in \cite{HeFokas}. In \cite{KAAN2013}, it was observed that NLS rogue patterns could be classified according to the order of the rogue waves and the parameter shifts applied to the Akhmediev breathers in the rogue-wave limit. This latter observation allowed the authors to extrapolate the shapes of rogue waves beyond order six, where numerical plotting of rogue waves became difficult. However, an analytical and quantitative prediction of NLS rogue patterns at arbitrary orders is still nonexistent.

In this article, we analytically investigate rogue wave patterns in the NLS equation at arbitrary rogue-wave orders. We show that if any internal parameter in the rogue waves (which controls the shape of initial small perturbations to the uniform background) is large, then these waves would exhibit clear geometric structures, which comprise Peregrine (fundamental) rogue waves forming shapes such as triangle, pentagon, heptagon and nonagon, with a possible lower-order rogue wave at the center. These rogue patterns are analytically predicted by the root structures of the Yablonskii-Vorob'ev polynomial hierarchy, and their orientations are controlled by the phase of the large internal parameter. We also show that if multiple internal parameters in the rogue waves are large but satisfy certain constraints, then the same rogue patterns would still hold. These results reveal a deep connection between rogue patterns and the Yablonskii-Vorob'ev polynomials, and drastically improve our analytical understanding and quantitative description of rogue events. As a small application of our analytical results, the numerical observation in \cite{KAAN2011} on single-shell ring structures is explained. Comparison between true rogue patterns and our analytical predictions is also presented, and excellent agreement is observed.

This paper is structured as follows. In Sec. 2, we present a simplified bilinear expression of general rogue waves in the NLS equation, as well as the Yablonskii-Vorob'ev polynomial hierarchy and its root structure, which we will utilize in later texts. In Sec. 3, we present our main theorems on rogue wave patterns when a single internal parameter in the rogue waves is large. In Sec. 4, we graphically illustrate rogue patterns under a large parameter and compare these patterns to our theoretical predictions. In Sec. 5, we prove the theorems stated in Sec. 3. In Sec. 6, we generalize our analytical results to cases where multiple internal parameters in the rogue waves are large but meet certain constraints. Sec. 7 concludes the paper.

\section{Preliminaries} \label{sec:pre}
The nonlinear Schr\"{o}dinger (NLS) equation
\[ \label{NLS-2020}
\textrm{i} u_{t} +  \frac{1}{2}u_{xx}+ |u|^2 u=0
\]
arises in numerous physical situations such as water waves and optics \cite{Benney,Ablowitzbook,Kivsharbook}. In this article, we consider its rogue wave solutions, which are rational solutions which approach a constant-amplitude continuous-wave background as $x, t\to \pm \infty$. Since this equation admits Galilean and scaling invariances, we can set the boundary conditions of these rogue waves as
\[ \label{bc}
u(x, t) \to e^{\textrm{i} t}, \quad x, t \to \infty,
\]
without any loss of generality.

\subsection{Improved bilinear expressions of general rogue waves} \label{sec:solution}
Analytical expressions for general rogue waves in the NLS equation have been derived in \cite{DGKM2010,GLML2012,OhtaJY2012} by various methods. However, those expressions are not the best for our solution analysis. Here, we present a simpler expression for these solutions, which can be derived by incorporating a new parameterization \cite{YangDNLS2019} into bilinear rogue waves of Ref.~\cite{OhtaJY2012}. These simpler expressions of rogue waves are given by the following theorem.
\begin{quote}
\textbf{Theorem 1.} The general NLS rogue waves under boundary conditions (\ref{bc}) are
\begin{eqnarray}
&& u_N(x,t)=\frac{\sigma_{1}}{\sigma_{0}}e^{\textrm{i}t}, \label{BilinearTrans2}
\end{eqnarray}
where the positive integer $N$ represents the order of the rogue wave, $\sigma_{n}$ is a $N \times N$ Gram determinant
\begin{equation} \label{sigma_n}
\sigma_{n}=
\det_{
\begin{subarray}{l}
1\leq i, j \leq N
\end{subarray}
}
\left(
\begin{array}{c}
 \phi_{2i-1,2j-1}^{(n)}
\end{array}
\right),
\end{equation}
the matrix elements in $\sigma_{n}$ are defined by
\begin{equation}
\phi_{i,j}^{(n)}=\sum_{\nu=0}^{\min(i,j)} \frac{1}{4^{\nu}} \hspace{0.06cm} S_{i-\nu}(\textbf{\emph{x}}^{+}(n) +\nu \textbf{\emph{s}})  \hspace{0.06cm} S_{j-\nu}(\textbf{\emph{x}}^{-}(n) + \nu \textbf{\emph{s}}),
\end{equation}
vectors $\textbf{\emph{x}}^{\pm}(n)=\left( x_{1}^{\pm}, x_{2}^{\pm},\cdots \right)$ are defined by
\begin{eqnarray} \label{xpmdef}
x_{1}^{\pm}=x \pm \textrm{i} t \pm n, \ \ \ x_{2k}^{\pm} = 0, \quad x_{2k+1}^{+}= \frac{x+2^{2k} (\textrm{i} t)}{(2k+1)!} +a_{2k+1},    \quad x_{2k+1}^{-}=  \frac{x-2^{2k} (\textrm{i} t)}{(2k+1)!}+ a_{2k+1}^*,
\end{eqnarray}
with $k\ge 1$ and the asterisk * representing complex conjugation,
$\textbf{\emph{s}}=(0, s_2, 0, s_4, \cdots)$ are coefficients from the expansion
\begin{eqnarray} \label{sexpand}
\sum_{j=1}^{\infty} s_{j}\lambda^{j}=\ln \left[\frac{2}{\lambda}  \tanh \left(\frac{\lambda}{2}\right)\right],
\end{eqnarray}
the Schur polynomials $S_k(\mbox{\boldmath $x$})$, with $\emph{\textbf{x}}=\left( x_{1}, x_{2}, \ldots \right)$, are defined by
\begin{equation} \label{Schurdef}
\sum_{k=0}^{\infty}S_k(\mbox{\boldmath $x$}) \epsilon^k
=\exp\left(\sum_{k=1}^{\infty}x_k \epsilon^k\right),
\end{equation}
or more explicitly,
\begin{equation} \label{Skdef}
S_{k}(\mbox{\boldmath $x$}) =\sum_{l_{1}+2l_{2}+\cdots+ml_{m}=k} \left( \ \prod _{j=1}^{m} \frac{x_{j}^{l_{j}}}{l_{j}!}\right),
\end{equation}
and $a_{2k+1} \hspace{0.05cm} (k=1, 2, \cdots, N-1)$ are free irreducible complex constants.
\end{quote}
This theorem will be proved in Appendix A. Since these rogue waves approach a uniform background as $t\to -\infty$, the internal parameters $\{a_{2k+1}\}$ in these waves physically control the shape of initial small perturbations to this uniform background, which in turn decide the subsequent time evolution and the resulting pattern of rogue waves.

As we will show, these rogue wave solutions will exhibit clear and recognizable patterns when some of these $N-1$ internal parameters $(a_3, a_5, \cdots, a_{2N-1})$ get large. It turns out that the resulting rogue patterns are determined by the root structures of the Yablonskii-Vorob'ev polynomial hierarchy, and this polynomial hierarchy and their root structures will be introduced next.

\subsection{The Yablonskii-Vorob'ev polynomial hierarchy and their root structures} \label{secPII}
Yablonskii-Vorob'ev polynomials arose in rational solutions of the second Painlev\'{e} equation ($\mbox{P}_{\mbox{\scriptsize II}}$) \cite{Yablonskii1959,Vorobev1965}
\[\label{PII}
w''=2 w^3+ z w+\alpha,
\]
where $\alpha$ is an arbitrary constant. It has been shown that this $\mbox{P}_{\mbox{\scriptsize II}}$ equation admits rational solutions if and only if $\alpha=N$ is an integer. In this case, the rational solution is unique and is given by
\begin{eqnarray}
&& w(z; N)= \frac{d}{dz} \ln\frac{Q_{N-1}(z)}{Q_{N}(z)}, \quad N\ge 1,   \label{wzn1}\\
&& w(z; 0)=0, \quad w(z; -N)=-w(z; N),      \label{wzn2}
\end{eqnarray}
and the polynomials $Q_N(z)$, now called the Yablonskii-Vorob'ev polynomials, are constructed by the following recurrence relation
\[
Q_{N+1}Q_{N-1}= z Q_{N}^2 -4 \left[ Q_{N}Q_{N}''-(Q_{N}')^2 \right],
\]
with $Q_{0}(z)=1$, $Q_{1}(z)=z$, and the prime denoting the derivative. Later, a determinant expression for these polynomials was found in \cite{Kajiwara-Ohta1996}. Let $p_{k}(z)$ be the special Schur polynomial defined by
\begin{equation}
\sum_{k=0}^{\infty}p_k(z) \epsilon^k =\exp\left( z \epsilon - \frac{4}{3}\epsilon^3 \right).
\end{equation}
Then, Yablonskii-Vorob'ev polynomials $Q_{N}(z)$ are given by the $N \times N$ determinant \cite{Kajiwara-Ohta1996}
\begin{eqnarray}
&& Q_{N}(z) = c_{N} \left| \begin{array}{cccc}
         p_{1}(z) & p_{0}(z) & \cdots &  p_{2-N}(z) \\
         p_{3}(z) & p_{2}(z) & \cdots &  p_{4-N}(z) \\
        \vdots& \vdots & \vdots & \vdots \\
         p_{2N-1}(z) & p_{2N-2}(z) & \cdots &  p_{N}(z)
       \end{array}
 \right|,
\end{eqnarray}
where $c_{N}= \prod_{j=1}^{N}(2j-1)!!$, and $p_{k}(z)= 0$ if $k<0$. These polynomials are monic polynomials with integer coefficients \cite{Clarkson2003-II}. The first few Yablonskii-Vorob'ev polynomials are
\begin{eqnarray*}
&& Q_2(z)=z^3+4, \\
&& Q_3(z)=z^6 + 20z^3 - 80,  \\
&& Q_4(z)=z(z^9 + 60z^6 + 11200).
\end{eqnarray*}

To define the Yablonskii-Vorob'ev polynomial hierarchy, we let $p_{k}^{[m]}(z)$ be the generalized Schur polynomial defined by
\begin{equation} \label{pkmz}
\sum_{k=0}^{\infty}p_k^{[m]}(z) \epsilon^k =\exp\left( z \epsilon - \frac{2^{2m}}{2m+1}\epsilon^{2m+1} \right),
\end{equation}
where $m$ is a positive integer. Then, the Yablonskii-Vorob'ev hierarchy $Q_{N}^{[m]}(z)$ are given by the $N \times N$ determinant \cite{Clarkson2003-II}
\begin{eqnarray} \label{QNm}
&& Q_{N}^{[m]}(z) = c_{N} \left| \begin{array}{cccc}
         p^{[m]}_{1}(z) & p^{[m]}_{0}(z) & \cdots &  p^{[m]}_{2-N}(z) \\
         p^{[m]}_{3}(z) & p^{[m]}_{2}(z) & \cdots &  p^{[m]}_{4-N}(z) \\
        \vdots& \vdots & \vdots & \vdots \\
         p^{[m]}_{2N-1}(z) & p^{[m]}_{2N-2}(z) & \cdots &  p^{[m]}_{N}(z)
       \end{array}
 \right|,
\end{eqnarray}
where $p_{k}^{[m]}(z)= 0$ if $k<0$. When $m=1$, $Q_{N}^{[1]}(z)$ are the original Yablonskii-Vorob'ev polynomials $Q_{N}(z)$. When $m>1$, $Q_{N}^{[m]}(z)$ give higher members of this polynomial hierarchy. All these $Q_{N}^{[m]}(z)$ polynomials were conjectured to be monic polynomials with integer coefficients as well \cite{Clarkson2003-II}. The first few $Q_{N}^{[2]}(z)$ polynomials are
\begin{eqnarray*}
&& Q_2^{[2]}(z)=z^3, \\
&& Q_3^{[2]}(z)=z(z^5-144),  \\
&& Q_4^{[2]}(z)=z^{10} - 1008 z^5-48384.
\end{eqnarray*}
These $Q_{N}^{[m]}(z)$ polynomials, through relations similar to (\ref{wzn1})-(\ref{wzn2}), provide the unique rational solution for the $\mbox{P}_{\mbox{\scriptsize II}}$ hierarchy \cite{Clarkson2003-II,Bertola2016}.
It is noted that the determinant (\ref{QNm}) for $Q_{N}^{[m]}(z)$ is a Wronskian, because it is easy to see from Eq.~(\ref{pkmz}) that
\[  \label{pkpkp1}
p_k^{[m]}(z)=[p_{k+1}^{[m]}]'(z).
\]

Root structures of the Yablonskii-Vorob'ev polynomial hierarchy have been studied before \cite{Fukutani,Taneda,Clarkson2003-II,Miller2014,Bertola2016}. Regarding the zero root, its multiplicity in $Q_{N}(z)$, $Q_{N}^{[2]}(z)$ and $Q_{N}^{[3]}(z)$ was presented in \cite{Taneda,Clarkson2003-II}. Generalizing those results, we have the following theorem.

\begin{quote}
\textbf{Theorem 2.} The general Yablonskii-Vorob'ev polynomial $Q_{N}^{[m]}(z)$ has degree $N(N+1)/2$, and is of the form
\[ \label{QNmform}
Q_{N}^{[m]}(z)=z^{N_0(N_0+1)/2}q_{N}^{[m]}(\zeta), \quad \zeta=z^{2m+1},
\]
where $q_{N}^{[m]}(\zeta)$ is a polynomial with a nonzero constant term, and the integer $N_0$ is given by the equation
\begin{eqnarray}
&&N\equiv N_{0} \hspace{0.1cm} \mbox{mod} \hspace{0.1cm} (2m+1), \quad \text{or}   \label{N01}\\
&&N\equiv -N_{0}-1 \hspace{0.1cm} \mbox{mod} \hspace{0.1cm} (2m+1),   \label{N02}
\end{eqnarray}
under the restriction of $0\leq  N_{0} \leq m$. Due to this restriction on $N_0$, only one of Eqs.  (\ref{N01}) and (\ref{N02}) can hold, and thus the resulting $N_0$ value is unique.
\end{quote}
The proof of this theorem will be provided in Appendix B. This theorem gives the multiplicity of the root zero in any $Q_{N}^{[m]}(z)$ polynomial. It also shows that the root structure of $Q_{N}^{[m]}(z)$ is invariant under $2\pi/(2m+1)$-angle rotation in the complex $z$ plane. In the particular case of the original Yablonskii-Vorob'ev polynomials $Q_{N}(z)$ where $m=1$, the above theorem shows that $0\le N_0\le 1$. This means that zero is either not a root or a simple root for $Q_{N}(z)$, in agreement with previous results in \cite{Fukutani,Taneda}.

On the determination of the unique $N_0$ value in the above theorem, let us give an example. When $N=5$ and $m=4$, the $N_0$ value under the restriction of $0\le N_0\le 4$ can only be found from Eq. (\ref{N02}) as $N_0=3$.

Regarding nonzero roots, it was shown in \cite{Fukutani} that for the original Yablonskii-Vorob'ev polynomials $Q_{N}(z)$, all nonzero roots are simple. For the higher Yablonskii-Vorob'ev polynomial hierarchy $Q_{N}^{[m]}(z)$, it was conjectured in \cite{Clarkson2003-II} that all nonzero roots are also simple.
In view of Theorem 2, this conjecture implies that the polynomial $Q_{N}^{[m]}(z)$ has
\[ \label{Np}
N_p=\frac{1}{2}\left[N(N+1)-N_{0}(N_{0}+1)\right]
\]
nonzero simple roots. We have checked this conjecture for a myriad of $(N, m)$ values and found it to hold in all our examples. Thus, we will assume it true and utilize it in our later analysis [see the sentence below Eq.~(\ref{sigmanxt5}) in Sec.~\ref{sec:proof}].

\begin{figure}[htb]
\begin{center}
\includegraphics[scale=0.85, bb=200 0 355 145]{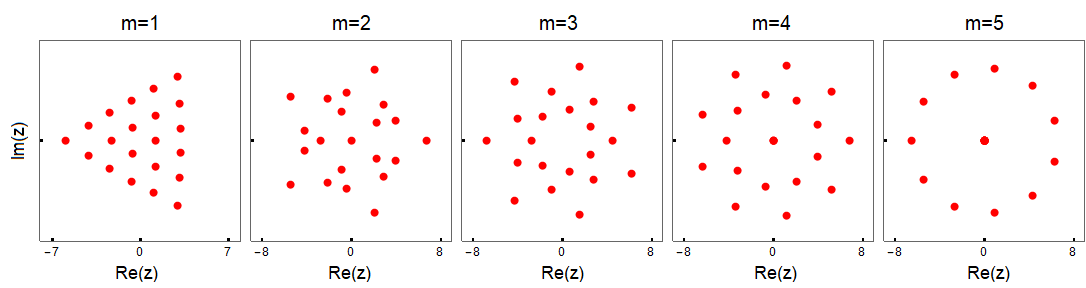}
\caption{Plots of the roots of the Yablonskii-Vorob'ev polynomial hierarchy $Q_{N}^{[m]}(z)$ for $N=6$ and $1\le m\le 5$.  } \label{f:roots}
\end{center}
\end{figure}

Roots of many Yablonskii-Vorob'ev polynomials $Q_{N}^{[m]}(z)$ were plotted in \cite{Clarkson2003-II}, and
highly regular and symmetric patterns were observed. Due to the importance of these root structures to our work, we reproduce some of those root plots in Fig.~\ref{f:roots} for $N=6$ and $1\le m\le 5$. Boundaries of the roots in  $Q_{N}^{[m]}(z)$ in the large-$N$ limit have been determined in \cite{Miller2014,Bertola2016}.

\section{Analytical predictions of rogue wave patterns for a single large internal parameter}
Rogue wave solutions in Theorem~1 contain $N-1$ free internal complex parameters $a_3, a_5, \cdots, a_{2N-1}$. In this section, we consider asymptotics of these rogue solutions when one of these internal parameters is large, while the other parameters remain $O(1)$. Generalizations to cases where multiple internal parameters are large but satisfy certain constraints will be made in Sec. \ref{sec:gen}.

Suppose $a_{2m+1}$ is large, where $1\le m\le N-1$, and the other parameters are $O(1)$. Then our results on the large-$a_{2m+1}$ asymptotics of rogue waves in Theorem~1 are summarized by the following two theorems.

\begin{quote}
\textbf{Theroem 3.}
Far away from the origin, with $\sqrt{x^2+t^2}=O\left(|a_{2m+1}|^{1/(2m+1)}\right)$, the $N$-th order rogue wave $u_N(x,t)$ in Eq. (\ref{BilinearTrans2}) separates into $N_p$ fundamental (Peregrine) rogue waves, where $N_p$ is given in Eq.~(\ref{Np}). These Peregrine waves are $\hat{u}_1(x-\hat{x}_{0}, t-\hat{t}_{0}) \hspace{0.05cm} e^{\textrm{i}t}$, where
\begin{equation} \label{Pere}
\hat{u}_1(x, t)=1- \frac{4(1+2\textrm{i}t)}{1+4x^2+4t^2},
\end{equation}
and their positions $(\hat{x}_{0}, \hat{t}_{0})$ are given by
\begin{eqnarray}
&&\hat{x}_{0}+\textrm{i}\hspace{0.05cm}\hat{t}_{0}=
z_{0}\left(-\frac{2m+1}{2^{2m}}a_{2m+1}\right)^{\frac{1}{2m+1}}, \label{x0t0}
\end{eqnarray}
with $z_{0}$ being any one of the $N_p$ simple nonzero roots of $Q_{N}^{[m]}(z)$. The error of this Peregrine wave approximation is $O(|a_{2m+1}|^{-1/(2m+1)})$. Expressed mathematically, when $\left[(x-\hat{x}_{0})^2+(t-\hat{t}_{0})^2\right]^{1/2}=O(1)$, we have the following solution asymptotics
\[ \label{Theorem3asym}
u_{N}(x,t; a_{3}, a_{5}, \cdots, a_{2N-1}) = \hat{u}_1(x-\hat{x}_{0},t-\hat{t}_{0})\hspace{0.05cm} e^{\textrm{i}t} + O\left(|a_{2m+1}|^{-1/(2m+1)}\right), \quad |a_{2m+1}|\gg 1.
\]
When $(x,t)$ is not in the neighborhood of any of these $N_p$ Peregrine waves, or $\sqrt{x^2+t^2}$ is larger than $O\left(|a_{2m+1}|^{1/(2m+1)}\right)$, $u_N(x,t)$ asymptotically approaches the constant background $e^{\textrm{i}t}$ as $|a_{2m+1}|\to \infty$.
\end{quote}

\begin{quote}
\textbf{Theroem 4.} In the neighborhood of the origin, where $\sqrt{x^2+t^2}=O(1)$, $u_N(x,t)$ is approximately a lower $N_0$-th order rogue wave $u_{N_0}(x,t)$, where $N_0$ is given in Theorem 2 with $0\le N_0\le N-2$, and $u_{N_0}(x,t)$ is given by Eq. (\ref{BilinearTrans2}) with its internal parameters $a_3, a_5, \cdots, a_{2N_0-1}$ being the first $N_0-1$ values in the parameter set $(a_3, a_5, \cdots, a_{2N-1})$ of the original rogue wave $u_N(x,t)$. The error of this lower-order rogue wave approximation $u_{N_0}(x,t)$ is $O(|a_{2m+1}|^{-1})$. Expressed mathematically, when $\sqrt{x^2+t^2}=O(1)$,
\[ \label{ucenter}
u_{N}(x,t; a_{3}, a_{5}, \cdots, a_{2N-1}) = u_{N_{0}}(x,t; a_{3}, a_{5}, \cdots, a_{2N_{0}-1})+ O\left(|a_{2m+1}|^{-1}\right), \quad |a_{2m+1}|\gg 1.
\]
If $N_0=0$, then there will not be such a lower-order rogue wave in the neighborhood of the origin, and $u_N(x,t)$ asymptotically approaches the constant background $e^{\textrm{i}t}$ there as $|a_{2m+1}|\to \infty$.
\end{quote}
These two theorems will be proved in Sec. \ref{sec:proof}.

\textbf{Remark 1.} Theorem 3 predicts that when $a_{2m+1}$ is large, the $N$-th order rogue wave (\ref{BilinearTrans2}) far away from the origin comprises $N_p$ Peregrine waves. The rogue pattern formed by these Peregrine waves has the same geometric shape as the root structure of the polynomial $Q_{N}^{[m]}(z)$, and thus this rogue pattern has $2\pi/(2m+1)$-angle rotational symmetry. The only difference between the predicted rogue pattern and the root structure of $Q_{N}^{[m]}(z)$ is a dilation and rotation between them due to the multiplication factor on the right side of Eq. (\ref{x0t0}). The angle of rotation is equal to the angle of the complex number $-a_{2m+1}$ divided by $2m+1$, and the dilation factor is equal to $[(2m+1)2^{-2m}|a_{2m+1}|]^{1/(2m+1)}$.

\textbf{Remark 2.} On the right side of Eq. (\ref{x0t0}), we can pick any one of the $(2m+1)$-th root of
$-(2m+1)2^{-2m}a_{2m+1}$, because roots $z_{0}$ of the polynomial $Q_{N}^{[m]}(z)$ have $2\pi/(2m+1)$-angle rotational symmetry, see the comment in the paragraph below Theorem 2.

As a small application of the above two theorems, we explain the numerical observations in Ref.~\cite{KAAN2011}. Under our bilinear rogue solution (\ref{BilinearTrans2}), a $N$-th order rogue wave exhibits a ring structure when $a_{2N-1}$ is large (see Fig. 2 in the next section). In this case, $m=N-1$, and $N_0=N-2$ from Eq.~(\ref{N02}). Then, our theory predicts that the center of this $N$-th order rogue wave is a $(N-2)$-th order rogue wave, surrounded by $N_p=2N-1$ Peregrine waves that are evenly spaced on a ring due to the $2\pi/(2m+1)$-, i.e., $2\pi/(2N-1)$-angle rotational symmetry (see Remark 1 above). This is precisely what was observed in~\cite{KAAN2011}.

\section{Comparison between true rogue patterns and our analytical predictions}
\begin{figure}[htb]
\begin{center}
\includegraphics[scale=0.8, bb=200 0 385 590]{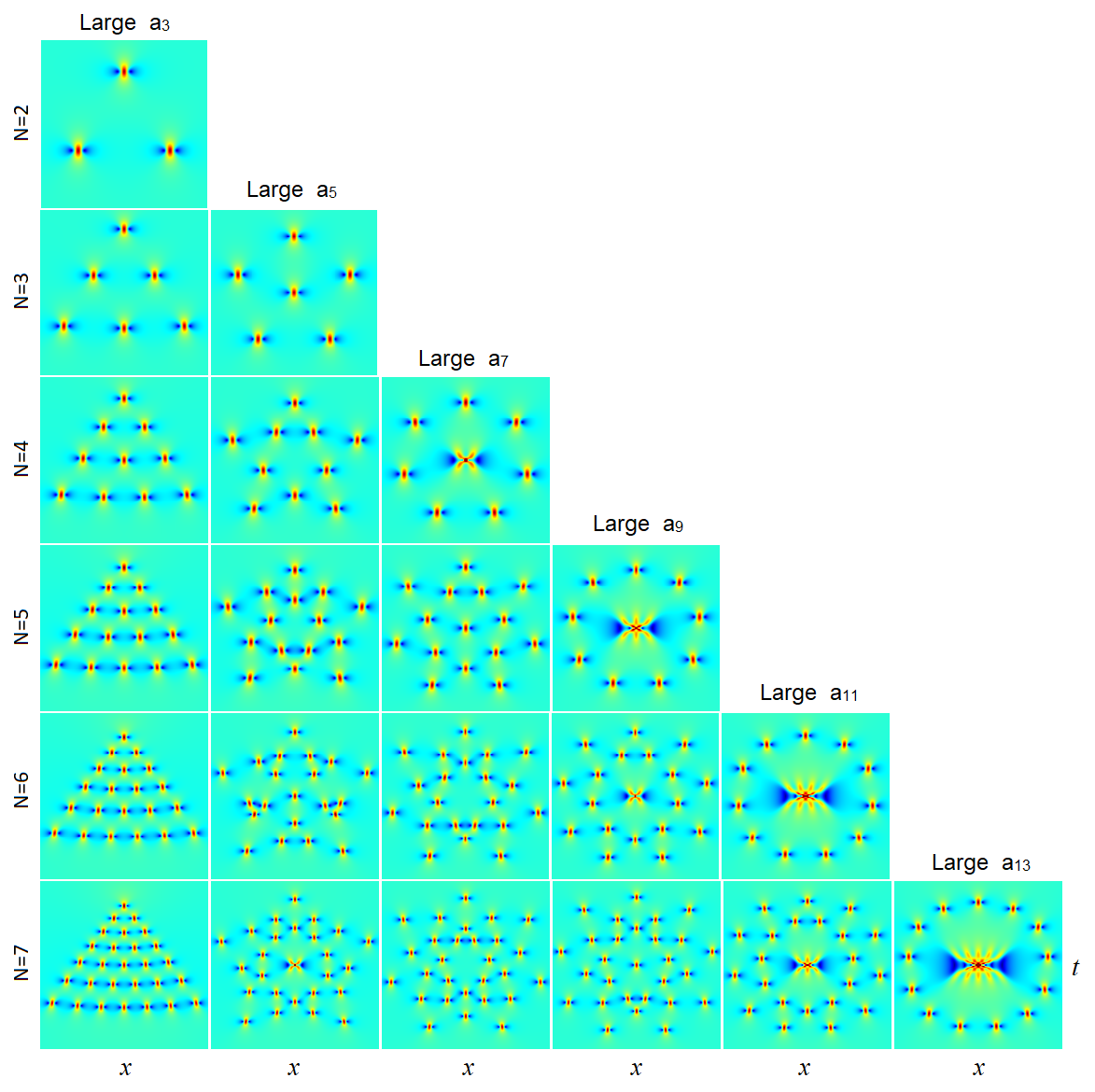}
\caption{True NLS rogue wave patterns $|u_N(x,t; a_{3}, a_{5}, \cdots, a_{2N-1})|$ from solutions (\ref{BilinearTrans2}) when $N$ ranges from $2$ to $7$ and one of the solution parameters is large (the other parameters are set as zero). The large parameter is labeled on top of each column, and its value for each panel is listed in Table 1. The center of each panel is always the origin $x=t=0$, but the $(x,t)$ intervals differ slightly from panel to panel. For instance, in the bottom row, the left-most panel has $-18.5 \le x, t \le 18.5$, and the right-most panel has $-16 \le x, t \le 16$. } \label{f:bigTrue}
\end{center}
\end{figure}

In this section, we compare true rogue patterns with our analytical predictions. For this purpose, we first show in Fig.~\ref{f:bigTrue} true rogue wave solutions (\ref{BilinearTrans2}) from the 2rd to 7th order, with large $a_3$, $a_5$, $a_7$, $a_9$, $a_{11}$ and $a_{13}$ in the first to sixth columns respectively. The specific value of the large parameter in each panel of this figure is listed in Table 1, and the other parameters in each solution are chosen as zero.

It is seen that these rogue waves comprise a number of Peregrine waves forming triangular patterns for large $a_3$, pentagon patterns for large $a_5$, heptagon patterns for large $a_7$, nonagon patterns for large $a_9$, hendecagon (eleven-sided polygon) patterns for large $a_{11}$, and tridecagon (thirteen-sided polygon) patterns for large $a_{13}$. In the literature, patterns on the diagonal of Fig.~\ref{f:bigTrue} are sometimes called single-shell ring structures \cite{KAAN2011}. In addition to these Peregrine waves away from the origin, some of the rogue waves also contain a lower-order rogue wave at their centers. For instance, for the 7-th order rogue waves in the bottom row of Fig.~\ref{f:bigTrue}, the first and fourth panels (with large $a_3$ and $a_9$ respectively) exhibit a Peregrine wave in their centers; the second panel (with large $a_5$) exhibits a second-order rogue wave in the center; the fifth panel (with large $a_{11}$) exhibits a third-order rogue wave in the center; and the last panel (with large $a_{13}$) exhibits a fifth-order rogue wave in the center. For our choices of parameters in rogue waves of Fig.~\ref{f:bigTrue}, these lower-order rogue waves in the center are all super-rogue waves, i.e., rogue waves with the highest peak amplitude of their orders. We note by passing that the first five rows of rogue patterns in Fig.~\ref{f:bigTrue} resemble those plotted in Ref.~\cite{KAAN2013} from Akhmediev breathers in the rogue-wave limit, although orientations between the two sets of patterns are very different.

\begin{table}[h]
\caption{Value of the large parameter for rogue waves in Fig.~\ref{f:bigTrue}}
\begin{center}
  \begin{tabular}{ | c | c | c|  c | c|  c | c|}  \hline
         $N$ &              $a_{3}$   & $a_{5}$    & $a_{7}$   & $a_{9}$   & $a_{11}$   & $a_{13}$   \\ \hline
          2 &               $-100 \textrm{i}$  &    &   &   &   &     \\ \hline
          3 &             $-60 \textrm{i}$  &  $-1000 \textrm{i}$ &     &   &   &    \\ \hline
          4 &              $-30 \textrm{i}$  &  $-300 \textrm{i}$ & $-3000 \textrm{i}$ &   &   &     \\ \hline
          5 &               $-20 \textrm{i}$  &  $-100 \textrm{i}$ & $-2000 \textrm{i}$ &$-12000 \textrm{i}$ & & \\ \hline
          6 &              $-20 \textrm{i}$  &  $-200 \textrm{i}$ & $ -2000 \textrm{i}$ &$-20000 \textrm{i}$ & $-80000 \textrm{i}$ &     \\ \hline
          7 &             $-20 \textrm{i}$  &  $-200 \textrm{i}$ & $ -2000 \textrm{i}$ &$-30000 \textrm{i}$ & $-100000 \textrm{i}$ & $-300000 \textrm{i}$   \\ \hline
  \end{tabular}
\end{center}
\end{table}

Now, we compare these true rogue patterns in Fig.~2 with our analytical predictions. Our prediction $|u_{N}^{(p)}(x,t)|$ from Theorems 3 and 4 can be assembled into a simple formula,
\begin{equation} \label{upNLS}
\left|u_{N}^{(p)}(x,t)\right|=\left|u_{N_{0}}(x,t)\right| + \sum _{j=1}^{N_{p}}  \left(\left| \hat{u}_1(x-\hat{x}_{0}^{(j)}, t-\hat{t}_{0}^{(j)})\right| -1 \right),
\end{equation}
where $\hat{u}_1(x,t)$ is the Peregrine wave given in (\ref{Pere}), their positions $(\hat{x}_{0}^{(j)}, \hat{t}_{0}^{(j)})$ given by (\ref{x0t0}) with $z_0$ being every one of the $N_p$ simple nonzero roots of $Q_{N}^{[m]}(z)$, and $u_{N_0}(x,t)$ is the lower-order rogue wave in Eq. (\ref{ucenter}) whose internal parameters $(a_{3}, a_{5}, \cdots, a_{2N_{0}-1})$ are the first $N_0-1$ values in the parameter set $(a_3, a_5, \cdots, a_{2N-1})$ of the original rogue wave $u_N(x,t)$. For true rogue waves in Fig.~\ref{f:bigTrue}, all internal parameters except for $a_{2m+1}$ were chosen as zero, and $N_0\le m$ (see Theorem 2). Then, all internal parameters in the predicted lower-order rogue wave $u_{N_{0}}(x,t)$ at the origin are also zero.

Our predicted $(N_p, N_0)$ values for rogue waves of Fig.~2 are displayed in Table 2, where $m=1, 2, \cdots, 6$ correspond to large $a_3, a_5, \cdots, a_{13}$ respectively. These $(N_p, N_0)$ values provide our predictions for the number of Peregrine waves away from the origin $(x,t)=(0,0)$, as well as the order of the reduced rogue wave in the neighborhood of the origin. Visual comparison between Table 2 and Fig. 2 shows complete agreement.

\begin{table}[h]
\caption{Predicted $(N_p, N_0)$ values for true rogue waves of Fig.~\ref{f:bigTrue}}
\begin{center}
  \begin{tabular}{ | l | l | l|  l | l|  l | l|}  \hline
         $N$ &              $m=1$   & $m=2$  & $m=3$   & $m=4$   & $m=5$   & $m=6$   \\ \hline
          2 &               (3, 0)  &    &    &   &      &     \\ \hline
          3 &               (6, 0)  & (5, 1)  &   &      &          &          \\ \hline
          4 &               (9, 1)  & (10, 0) & (7, 2)   &          &        &     \\ \hline
          5 &               (15, 0) & (15, 0) & (14, 1)  &  (9, 3)  &        & \\ \hline
          6 &               (21, 0) & (20, 1) & (21, 0)  &  (18, 2) &  (11, 4)&     \\ \hline
          7 &               (27, 1) & (25, 2) & (28, 0)  &  (27, 1) &  (22, 3)&  (13, 5)  \\ \hline
  \end{tabular}
\end{center}
\end{table}

\begin{figure}[htb]
\begin{center}
\includegraphics[scale=0.8, bb=200 0 385 580]{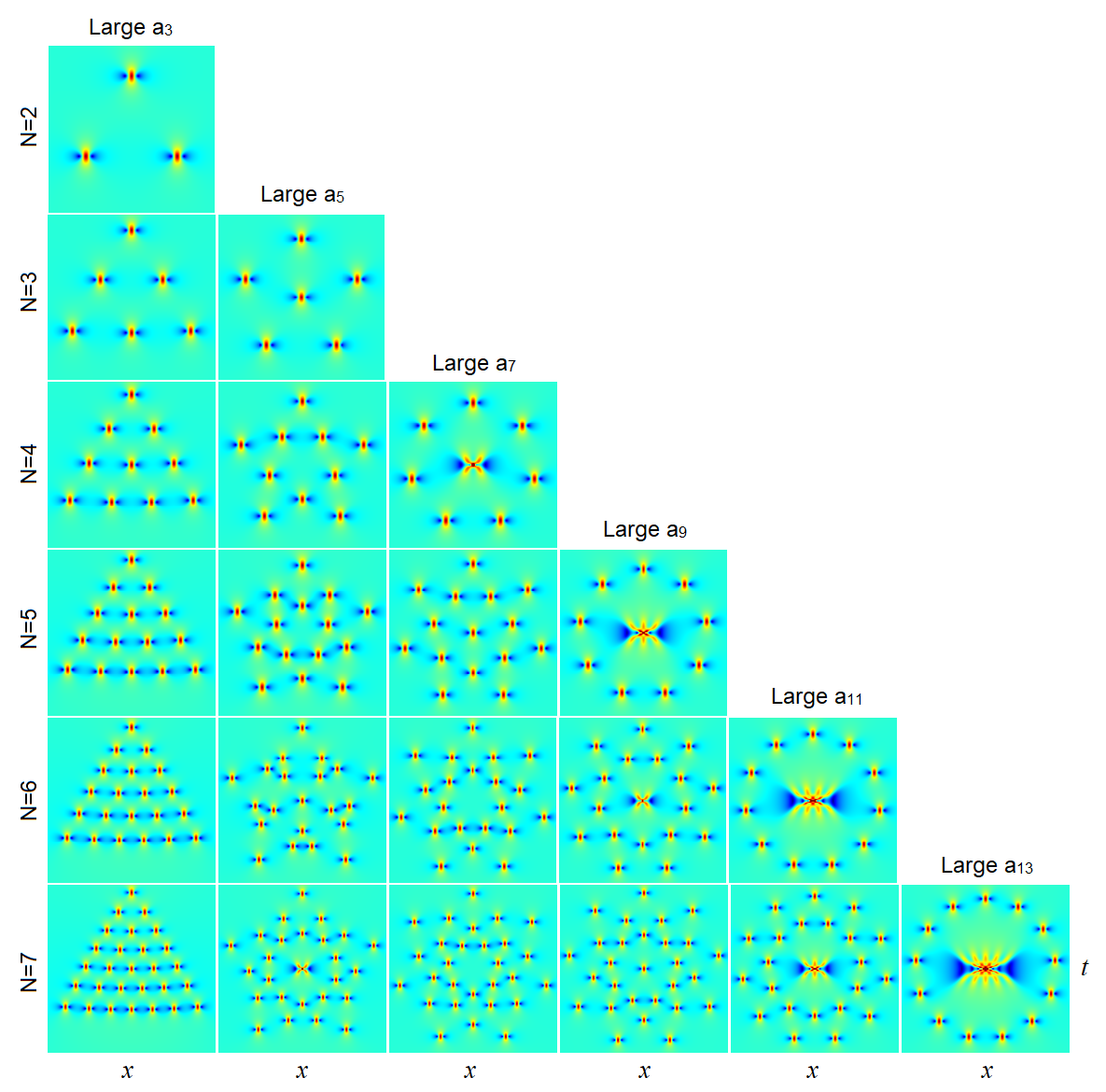}
\caption{Analytical predictions (\ref{upNLS}) for true rogue waves in Fig.~\ref{f:bigTrue}. The $x$ and $t$ intervals here are identical to those in Fig.~\ref{f:bigTrue}. }  \label{f:bigPred}
\end{center}
\end{figure}

We further compare our predicted whole solutions (\ref{upNLS}) with the true solutions of Fig.~\ref{f:bigTrue}
for the same sets of $(a_{3}, a_{5}, \cdots)$ parameter values. These predicted whole solutions (\ref{upNLS}) are displayed in Fig.~\ref{f:bigPred}, with identical $(x,t)$ intervals as in Fig.~\ref{f:bigTrue}'s true solutions. It is seen that the predicted patterns are strikingly similar to the true ones. In particular, since our predicted Peregrine locations (\ref{x0t0}) in the $(x, t)$ plane are given by all the non-zero roots of the Yablonskii-Vorob'ev polynomials $Q_{N}^{[m]}(z)$, multiplied by a fixed complex constant, predicted patterns formed by these Peregrine waves then are simply the root structures of these Yablonskii-Vorob'ev polynomials under certain rotation and dilation, as is evident by comparing predicted rogue waves in Fig.~\ref{f:bigPred} to the Yablonskii-Vorob'ev root structures in Fig.~\ref{f:roots} for $N=6$. These predicted Peregrine patterns clearly match the true ones in Fig.~\ref{f:bigTrue} very well. This visual agreement shows the deep connection between NLS rogue patterns and root structures of the Yablonskii-Vorob'ev hierarchy, as our theorem 3 predicts.

Regarding our predictions $u_{N_{0}}(x,t)$ for centers of the rogue waves $u_N(x,t)$ in Fig.~\ref{f:bigTrue}, we can show that our bilinear rogue wave solution (\ref{BilinearTrans2}) in Theorem 1 with all internal parameters set as zero gives the super-rogue wave. This means that our predictions $u_{N_{0}}(x,t)$ for the centers of true rogue waves are all lower-order super-rogue waves, which agree with centers of true solutions shown in Fig.~\ref{f:bigTrue}.

\begin{figure}[htb]
\begin{center}
\includegraphics[scale=0.350, bb=500 0 385 350]{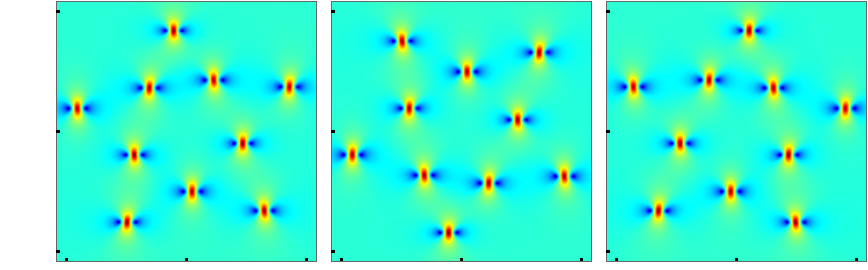}

\includegraphics[scale=0.350, bb=500 0 385 350]{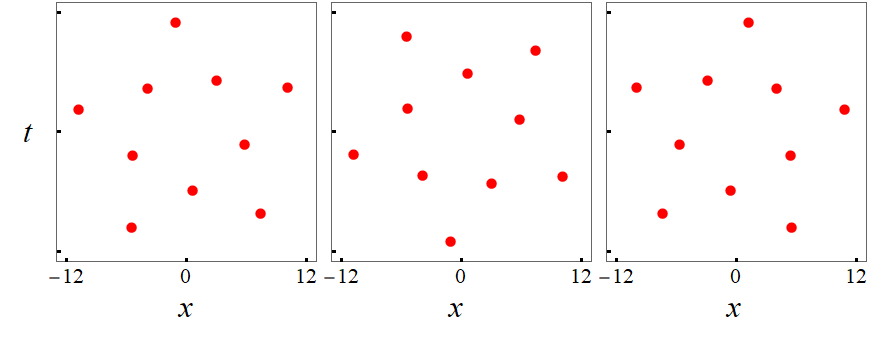}
\caption{Orientations of 4-th order pentagon-shaped rogue waves with $a_5=500 e^{-\textrm{i} \pi/3}$ (left column), $500 e^{\textrm{i} \pi/3}$ (middle column) and $500 e^{4\textrm{i} \pi/3}$ (right column) respectively; the other parameters in the rogue solutions are zero. Upper row: true rogue patterns from solutions (\ref{BilinearTrans2}); lower row: predicted locations of Peregrine waves from Eq. (\ref{x0t0}). } \label{f:orien}
\end{center}
\end{figure}

Theorem 3 reveals that the orientation of the rogue pattern formed by Peregrine waves is controlled by the phase of the large parameter $a_{2m+1}$. Specifically, the rogue-pattern orientation is the one of the root pattern of $Q_{N}^{[m]}(z)$ rotated by an angle of $\mbox{arg}(-a_{2m+1})/(2m+1)$, where ``arg" represents the argument (angle) of a complex number. To check this prediction, we choose the 4-th order pentagon-shaped rogue waves, where $a_5$ is large and the other parameters are set as zero. For three choices of the $a_5$ value with the same modulus but different arguments, namely, $500 e^{-\textrm{i} \pi/3}$, $500 e^{\textrm{i} \pi/3}$ and $500 e^{4\textrm{i} \pi/3}$, true rogue patterns from solutions (\ref{BilinearTrans2}) are displayed in the upper row of Fig.~\ref{f:orien}. As expected, orientations of these pentagon patterns indeed change as the argument of $a_5$ varies. Using our formula (\ref{x0t0}), predicted locations of Peregrine waves in the rogue pattern are shown in the lower row of Fig.~\ref{f:orien}. Comparison of the upper and lower rows of Fig.~\ref{f:orien} shows that the predicted orientations are in perfect agreement with the true ones.

Next, we make quantitative comparisons between true rogue waves and our predictions for large $a_{2m+1}$, and verify the error decay rate of $O(|a_{2m+1}|^{-1/(2m+1)})$ for the prediction of Peregrine-wave locations far away from the origin in Theorem 3, and the error decay rate of $O(|a_{2m+1}|^{-1})$ for the prediction of the lower-order rogue wave at the center in Theorem 4.

\begin{figure}[htb]
\begin{center}
\includegraphics[scale=0.8, bb=200 0 385 170]{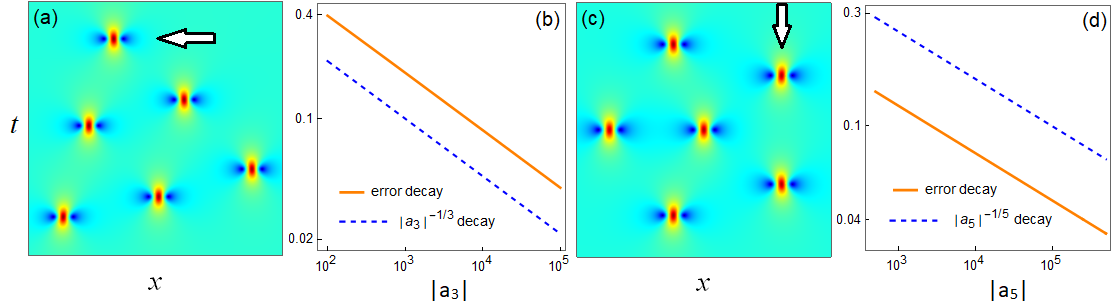}
\caption{Decay of errors in our prediction (\ref{x0t0}) for the Peregrine location as $|a_3|$ or $|a_5|$ increases. (a) A triangle pattern of 3rd-order rogue waves when $|a_{3}|$ is large and arg$(a_3)=-\pi /4$. (b) Error versus $|a_3|$ for the Peregrine location marked by an arrow in (a).
(c) A pentagon pattern of 3rd-order rogue waves when $|a_5|$ is large with arg$(a_5)=-\pi /4$. (b) Error versus $|a_5|$ for the Peregrine location marked by an arrow in (c).}   \label{f:comPere}
\end{center}
\end{figure}

For the quantitative comparison on Peregrine-wave locations away from the origin, we choose two patterns of 3rd-order rogue waves. One is a triangle pattern from large $a_3$, and we set arg$(a_3)=-\pi /4$; and the other is a pentagon pattern from large $a_5$, and we set $a_5$ to be real positive. In each pattern, we choose all other parameters of the rogue wave solutions to be zero. These triangul and pentagon patterns are shown schematically in Fig.~\ref{f:comPere}(a, c) respectively. In each of these two patterns, we pick one of its Peregrine waves, which is marked by an arrow, and quantitatively compare its true $(x_0, t_0)$ location with our analytical prediction (\ref{x0t0}) as $|a_3|$ or $|a_5|$ increases. Here, the true location of the Peregrine wave is defined as the $(x_0, t_0)$ location where this Peregrine wave attains its maximum amplitude, and the error of our asymptotic prediction $(\hat{x}_{0},\ \hat{t}_{0})$ in Eq.~(\ref{x0t0}) is defined as
\begin{equation*}
\mbox{error of Peregrine location} = \sqrt{\left(\hat{x}_{0}-x_{0} \right)^2+\left(\hat{t}_{0}-t_{0}\right)^2}.
\end{equation*}
These errors of Peregrine locations versus $|a_3|$ or $|a_5|$ are plotted as solid lines in panels (b) and (d) of Fig.~\ref{f:comPere} for the triangular and pentagon patterns respectively. For comparison, the decay rates of $|a_{3}|^{-1/3}$ and $|a_{5}|^{-1/5}$ are also displayed in these panels as dashed lines. We see that these errors of Peregrine locations indeed decay at the rate of $|a_{2m+1}|^{-1/(2m+1)}$, thus confirming the analytical error estimates (\ref{Theorem3asym}) in Theorem~3.

\begin{figure}[htb]
\begin{center}
\includegraphics[scale=0.8, bb=200 0 385 170]{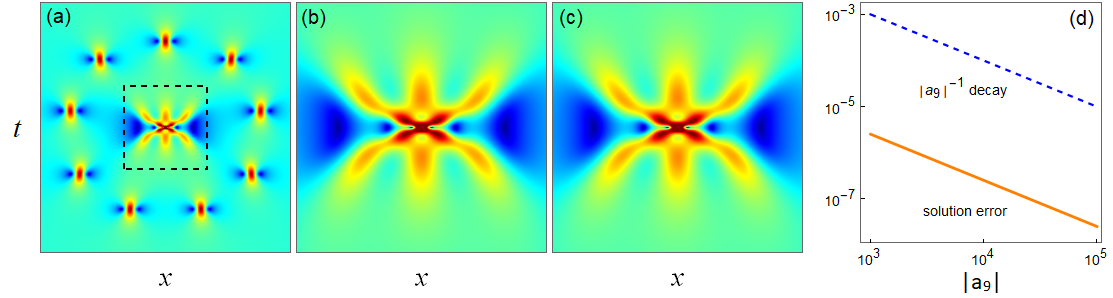}
\caption{Decay of errors in our prediction $u_{3}(x,t)$ for the center region of the rogue wave $u_{5}(x,t)$ with large $a_{9}$. (a) A true 5-th order rogue wave $|u_{5}(x,t)|$ with $a_{9}=-5000 \textrm{i}$ and the other parameters being zero; the $(x,t)$ intervals here are $-12\le x, t \le 12$. (b) Zoomed-in plot of the center region of the true solution marked by a dashed-line box in panel (a). (c) Our prediction $|u_{3}(x,t)|$ for the center region with the same $(x,t)$ intervals as in (b). (d) Error decay of our predicted solution at the $(x,t)$ location of $(0.5, 0.5)$ as $a_9$ increases in size with arg$(a_9)=-\pi /2$.  } \label{f:comcenter}
\end{center}
\end{figure}

To quantitatively compare our prediction in Theorem~4 on the lower-order rogue wave at the center with the true solution, we choose a fifth-order rogue wave $u_{5}(x,t)$ with large $a_{9}$ and the other internal parameters set as zero. This $|u_5(x,t)|$ solution with $a_{9}=-5000 \textrm{i}$ is displayed in Fig.~\ref{f:comcenter}(a). The center region of this wave marked by a dashed-line box in panel (a) is amplified and replotted in panel (b). In the present case, $N=5$ and $m=4$. Since $5\equiv -4 \hspace{0.1cm} \mbox{mod} \hspace{0.1cm} 9$, we get $N_0=3$ from Eq.~(\ref{N02}). Thus, according to Theorem 4, this $u_5(x,t)$ solution contains a 3rd-order rogue wave $u_3(x,t)$ in its center region, where all internal parameters $(a_3, a_5)$ in this $u_3(x,t)$ solution are zero. Such a $u_3(x,t)$ solution is a third-order super rogue wave. This predicted $|u_3(x,t)|$ solution is displayed in Fig.~\ref{f:comcenter}(c), with the same $(x,t)$ internals as in the true center-region solution displayed in panel (b). Visually, this predicted center solution in (c) is identical to the true center solution in (b). Quantitatively, we have also obtained the errors in our predicted solution $u_{3}(x,t)$ at $x=t=0.5$ of the center region as $a_9$ increases in magnitude with arg$(a_9)=-\pi /2$. Our error is defined as
\begin{equation*}
\mbox{error of center region prediction} = \left|u_5(x,t)-u_3(x,t)\right|_{x=t=0.5}.
\end{equation*}
The dependence of this error on $|a_9|$ is plotted in Fig.~\ref{f:comcenter}(d). Comparison of this error decay with the $|a_9|^{-1}$ decay [shown as a dashed line in panel (d)] indicates that this error is indeed of $O(|a_{9}|^{-1})$, confirming the error prediction (\ref{ucenter}) in Theorem~4.

\section{Proofs for the analytical results in Sec. 3} \label{sec:proof}
In this section, we prove the analytical predictions on NLS rogue patterns in Theorems 3 and 4 of Sec. 3. Our proof is based on an asymptotic analysis of the rogue wave solution (\ref{BilinearTrans2}), or equivalently, the determinant $\sigma_n$ in Eq. (\ref{sigma_n}), in the large $a_{2m+1}$ limit.

\textbf{Proof of Theorem 3.} \hspace{0.05cm}
When $a_{2m+1}$ is large and the other parameters $O(1)$ in the rogue wave solution (\ref{BilinearTrans2}),
at $(x,t)$ where $\sqrt{x^2+t^2}=O\left(|a_{2m+1}|^{1/(2m+1)}\right)$, by denoting
\[ \label{deflambda}
\lambda=a_{2m+1}^{-1/(2m+1)}
\]
and recalling the expression of Schur polynomials in Eq.~(\ref{Skdef}), we have
\begin{eqnarray}\label{Skasym1}
&& S_{k}(\textbf{\emph{x}}^{+}(n) +\nu \textbf{\emph{s}}) =
S_{k}\left(x_{1}^{+}, \nu s_{2}, x_{3}^{+}, \nu s_4, \cdots\right)=\lambda^{-k} S_{k}\left(x_{1}^{+}\lambda, \nu s_{2} \lambda^2, x_{3}^{+} \lambda^3, \nu s_{4}\lambda^4, \cdots\right) \nonumber \\
&&\sim \lambda^{-k} S_{k}\left[(x+\textrm{i}t)\lambda, 0, \cdots, 0, 1, 0, \cdots\right]
=S_{k}\left(x+\textrm{i}t, 0, \cdots, 0, a_{2m+1}, 0, \cdots\right).
\end{eqnarray}
Thus,
\[ \label{Skxnasym}
S_k(\textbf{\emph{x}}^{+}(n) +\nu \textbf{\emph{s}})  \sim S_k(\textbf{v}), \quad \quad |a_{2m+1}| \gg 1,
\]
where
\[  \label{vdef}
\textbf{v}=(x+\textrm{i}t, 0, \cdots, 0, a_{2m+1}, 0, \cdots).
\]
From the definition of Schur polynomials (\ref{Schurdef}), $S_k(\textbf{v})$ is given by
\begin{equation}
\sum_{k=0}^{\infty}S_k(\textbf{v}) \epsilon^k
=\exp\left[(x+\textrm{i}t)\epsilon +a_{2m+1}\epsilon^{2m+1}\right].
\end{equation}
Thus, it is related to the polynomial $p_{k}^{[m]}(z)$ in (\ref{pkmz}) as
\begin{equation} \label{Skorder}
S_k(\textbf{v})=A^{k/(2m+1)}p_{k}^{[m]}(z),
\end{equation}
where
\begin{equation} \label{Az}
A=-\frac{2m+1}{2^{2m}}a_{2m+1}, \hspace{0.3cm} z=A^{-1/(2m+1)}(x+\textrm{i}t).
\end{equation}
Using these formulae, we find that
\begin{equation} \label{Splusasym}
\det_{1 \leq i, j\leq N} \left[S_{2i-j}(\textbf{\emph{x}}^{+}(n) +j \textbf{\emph{s}}) \right]
\sim c_N^{-1}A^{\frac{N(N+1)}{2(2m+1)}}Q_{N}^{[m]}(z), \quad \quad |a_{2m+1}| \gg 1.
\end{equation}
Similarly,
\begin{equation} \label{Sminusasym}
\det_{1 \leq i, j\leq N} \left[S_{2i-j}(\textbf{\emph{x}}^{-}(n) +j \textbf{\emph{s}}) \right] \sim
c_N^{-1}\left(A^*\right)^{\frac{N(N+1)}{2(2m+1)}}Q_{N}^{[m]}(z^*), \quad \quad |a_{2m+1}| \gg 1.
\end{equation}
Hereafter, $S_k= 0$ when $k<0$.

To proceed further, we use determinant identities and the Laplace expansion to rewrite $\sigma_n$ in Eq. (\ref{sigma_n}) as \cite{OhtaJY2012}
\begin{eqnarray} \label{sigmanLap}
&& \hspace{-0.8cm} \sigma_{n}=\sum_{0\leq\nu_{1} < \nu_{2} < \cdots < \nu_{N}\leq 2N-1}
\det_{1 \leq i, j\leq N} \left[\frac{1}{2^{\nu_j}} S_{2i-1-\nu_j}(\textbf{\emph{x}}^{+}(n) +\nu_j \textbf{\emph{s}}) \right]  \times \det_{1 \leq i, j\leq N}\left[\frac{1}{2^{\nu_j}}S_{2i-1-\nu_j}(\textbf{\emph{x}}^{-}(n) + \nu_j \textbf{\emph{s}})\right].
\end{eqnarray}
Since the highest order term of $a_{2m+1}$ in this $\sigma_n$ comes from the index choices of $\nu_{j}=j-1$, then
\begin{equation} \label{sigmanmax}
\sigma_n \sim |\alpha|^2 \hspace{0.05cm} |a_{2m+1}|^{\frac{N(N+1)}{2m+1}} \left|Q_{N}^{[m]}(z)\right|^2,
\quad \quad |a_{2m+1}| \gg 1,
\end{equation}
where
\[ \label{defalpha}
\alpha=2^{-N(N-1)/2}c_N^{-1} \left(-\frac{2m+1}{2^{2m}}\right)^{\frac{N(N+1)}{2(2m+1)}}.
\]
Since $\alpha$ is independent of $n$, the above equation shows that for large $a_{2m+1}$, $\sigma_1/\sigma_0\sim 1$, i.e., the solution $u(x,t)$ is on the unit background, except at or near $(x, t)$ locations $\left(\hat{x}_{0}, \hat{t}_{0}\right)$ where
\[  \label{z0def}
z_0=A^{-1/(2m+1)}(\hat{x}_{0}+\textrm{i}\hat{t}_{0})
\]
is a root of the polynomial $Q_{N}^{[m]}(z)$, and such $\left(\hat{x}_{0}, \hat{t}_{0}\right)$ locations are given by Eq. (\ref{x0t0}) in view of Eq. (\ref{Az}).

Next, we show that when $(x, t)$ is in the neighborhood of each of the $\left(\hat{x}_{0}, \hat{t}_{0}\right)$ locations given by Eq. (\ref{x0t0}), i.e., when $\left[(x-\hat{x}_{0})^2+(t-\hat{t}_{0})^2\right]^{1/2}=O(1)$, the rogue wave $u_N(x,t)$ in Eq. (\ref{BilinearTrans2}) approaches a Peregrine wave $\hat{u}_1(x-\hat{x}_{0}, t-\hat{t}_{0})\hspace{0.05cm} e^{\textrm{i}t}$ for large $a_{2m+1}$. The asymptotic analysis above indicates that when $(x, t)$ is in the neighborhood of $\left(\hat{x}_{0}, \hat{t}_{0}\right)$, the highest power term $|a_{2m+1}|^{\frac{N(N+1)}{2m+1}}$ in $\sigma(x,t)$ vanishes. Thus, in order to determine the asymptotics of $u_N(x,t)$ in that $(x,t)$ region, we need to derive the leading order term of $a_{2m+1}$ in Eq. (\ref{sigmanLap}) whose order is lower than $|a_{2m+1}|^{\frac{N(N+1)}{2m+1}}$. For this purpose, we notice from Eq.~(\ref{Skasym1}) that when $(x, t)$ is in the neighborhood of $\left(\hat{x}_{0}, \hat{t}_{0}\right)$, we have a more refined asymptotics for $S_k(\textbf{\emph{x}}^{+}(n)+\nu \textbf{\emph{s}})$ as
\begin{eqnarray}\label{Skasym2}
&& S_{k}(\textbf{\emph{x}}^{+}(n) +\nu \textbf{\emph{s}}) = \lambda^{-k} S_{k}\left(x_1^+ \lambda, 0, \cdots, 0, 1, 0, \cdots\right)\left[1+O(\lambda^{2})\right] \nonumber \\
&&\hspace{2.1cm} =S_{k}\left(x_1^+, 0, \cdots, 0, a_{2m+1}, 0, \cdots\right)\left[1+O(\lambda^{2})\right],
\end{eqnarray}
i.e.,
\[ \label{Skxnasym2}
S_k(\textbf{\emph{x}}^{+}(n) +\nu \textbf{\emph{s}}) = S_k(\hat{\textbf{v}}) \left[1+O\left(a_{2m+1}^{-2/(2m+1)}\right)\right],
\]
where
\[\label{vhatdef}
\hat{\textbf{v}}=(x+\textrm{i}t+n, 0, \cdots, 0, a_{2m+1}, 0, \cdots).
\]
The polynomials $S_k(\hat{\textbf{v}})$ are related to $p_{k}^{[m]}(z)$ in (\ref{pkmz}) as
\begin{equation} \label{Skorder2}
S_k(\hat{\textbf{v}})=A^{k/(2m+1)}p_{k}^{[m]}(\hat{z}),
\end{equation}
where $A$ is as given in Eq.~(\ref{Az}), and $\hat{z}=A^{-1/(2m+1)}(x+\textrm{i}t+n)$.

Now, we derive the leading order term of $a_{2m+1}$ in Eq. (\ref{sigmanLap}). This leading order term comes from two index choices, one being $\nu=(0, 1, \cdots, N-1)$, and the other being $\nu=(0, 1, \cdots, N-2, N)$.

With the first index choice, in view of Eqs. (\ref{Skxnasym2}) and (\ref{Skorder2}), the determinant involving $\textbf{\emph{x}}^{+}(n)$ in Eq.~(\ref{sigmanLap}) is
\[ \label{Phinnxtnew}
\alpha \hspace{0.06cm} a_{2m+1}^{\frac{N(N+1)}{2(2m+1)}}Q_{N}^{[m]}(\hat{z}) \left[1+O\left(a_{2m+1}^{-2/(2m+1)}\right)\right],
\]
where $\alpha$ is given in Eq.~(\ref{defalpha}). Expanding $Q_{N}^{[m]}(\hat{z})$ around $\hat{z}=z_0$, where $z_0$ is given in Eq.~(\ref{z0def}), and recalling $Q_{N}^{[m]}(z_0)=0$, we have
\[
Q_{N}^{[m]}(\hat{z})=A^{-1/(2m+1)}\left[(x-\hat{x}_{0})+\textrm{i}(t-\hat{t}_{0})+n\right]\left[Q_{N}^{[m]}\right]'(z_0)
\left[1+O\left(A^{-1/(2m+1)}\right)\right].
\]
Inserting this equation into (\ref{Phinnxtnew}) and recalling the definition of $A$ in (\ref{Az}), the determinant involving $\textbf{\emph{x}}^{+}(n)$ in Eq.~(\ref{sigmanLap}) becomes
\[
\left[(x-\hat{x}_{0})+\textrm{i}(t-\hat{t}_{0})+n\right] \hspace{0.06cm} \hat{\alpha} \hspace{0.06cm} a_{2m+1}^{\frac{N(N+1)-2}{2(2m+1)}}\left[Q_{N}^{[m]}\right]'(z_0)\left[1+O\left(a_{2m+1}^{-1/(2m+1)}\right)\right],
\]
where $\hat{\alpha}=\alpha \hspace{0.04cm} [-(2m+1)2^{-2m}]^{-1/(2m+1)}$. Similarly, the
determinant involving $\textbf{\emph{x}}^{-}(n)$ in Eq.~(\ref{sigmanLap}) becomes
\[
\left[(x-\hat{x}_{0})-\textrm{i}(t-\hat{t}_{0})-n\right] \hspace{0.06cm} \hat{\alpha}^* \hspace{0.06cm} (a_{2m+1}^*)^{\frac{N(N+1)-2}{2(2m+1)}}\left[Q_{N}^{[m]}\right]'(z_0^*)\left[1+O\left(a_{2m+1}^{-1/(2m+1)}\right)\right].
\]

Next, we consider the contribution in Eq.~(\ref{sigmanLap}) from the second index choice of $\nu=(0, 1, \cdots, N-2, N)$. For this index choice, the determinant involving $\textbf{\emph{x}}^{+}(n)$ in Eq.~(\ref{sigmanLap}) is
\[ \label{sigmadet1}
\det_{1 \leq i \leq N} \left[S_{2i-1}(\textbf{\emph{x}}^{+}), \frac{1}{2}S_{2i-2}(\textbf{\emph{x}}^{+} + \textbf{\emph{s}}), \cdots, \frac{1}{2^{N-2}}S_{2i-(N-1)}[\textbf{\emph{x}}^{+} +(N-2) \hspace{0.06cm} \textbf{\emph{s}}], \frac{1}{2^{N}}S_{2i-(N+1)}(\textbf{\emph{x}}^{+} +N \hspace{0.04cm}\textbf{\emph{s}})\right].
\]
Utilizing Eqs. (\ref{Skxnasym2})-(\ref{Skorder2}), this determinant is
\[
2^{-N(N-1)/2-1}A^{\frac{N(N+1)-2}{2(2m+1)}}
\det_{1 \leq i \leq N} \left[p^{[m]}_{2i-1}(\hat{z}),  p^{[m]}_{2i-2}(\hat{z}),
\cdots, p^{[m]}_{2i-(N-1)}(\hat{z}),  p^{[m]}_{2i-(N+1)}(\hat{z})\right]\left[1+O\left(a_{2m+1}^{-2/(2m+1)}\right)\right].
\]
Recalling Eq. (\ref{pkpkp1}), we see that $p^{[m]}_{2i-(N+1)}(\hat{z})=[p^{[m]}_{2i-N}]'(\hat{z})$. Thus, the determinant in the above expression is equal to $c_N^{-1}\left[Q_{N}^{[m]}\right]'(\hat{z})$, so that the determinant (\ref{sigmadet1}) becomes
\[ \label{Phinnxtnew2}
\frac{1}{2}\hat{\alpha} \hspace{0.06cm} a_{2m+1}^{\frac{N(N+1)-2}{2(2m+1)}}\left[Q_{N}^{[m]}\right]'(\hat{z}) \left[1+O\left(a_{2m+1}^{-2/(2m+1)}\right)\right].
\]
When $(x, t)$ is in the neighborhood of $(\hat{x}_{0}, \hat{t}_{0})$, we expand $\left[Q_{N}^{[m]}\right]'(\hat{z})$ around $\hat{z}=z_0$ to reduce this expression further to
\[
\frac{1}{2}\hat{\alpha} \hspace{0.06cm} a_{2m+1}^{\frac{N(N+1)-2}{2(2m+1)}}\left[Q_{N}^{[m]}\right]'(z_0) \left[1+O\left(a_{2m+1}^{-1/(2m+1)}\right)\right].
\]
Similarly, the determinant involving $\textbf{\emph{x}}^{-}(n)$ in Eq.~(\ref{sigmanLap}) becomes
\[
\frac{1}{2}\hat{\alpha}^* \hspace{0.06cm} (a_{2m+1}^*)^{\frac{N(N+1)-2}{2(2m+1)}}\left[Q_{N}^{[m]}\right]'(z_0^*) \left[1+O\left(a_{2m+1}^{-1/(2m+1)}\right)\right].
\]

Summarizing the above two contributions, we find that
\[ \label{sigmanxt5}
\sigma_{n}(x,t) = |\hat{\alpha}|^2 \hspace{0.06cm} \left|\left[Q_{N}^{[m]}\right]'(z_0)\right|^2 |a_{2m+1}|^{\frac{N(N+1)-2}{(2m+1)}}
\left[ \left(x-\hat{x}_{0}\right)^2+\left(t-\hat{t}_{0}\right)^2 - (2 \textrm{i}) n \left(t-\hat{t}_{0}\right)-n^2+\frac{1}{4} \right]\left[1+O\left(a_{2m+1}^{-1/(2m+1)}\right)\right].
\]
Finally, we recall that nonzero roots are simple in Yablonskii-Vorob'ev polynomials $Q_{N}(z)$ \cite{Fukutani}. In addition, nonzero roots have also been conjectured to be simple in all the Yablonskii-Vorob'ev hierarchy $Q_{N}^{[m]}(z)$ \cite{Clarkson2003-II}. Assuming this conjecture is true, then
$\left[Q_{N}^{[m]}\right]'(z_0)\ne 0$. This indicates that the above leading-order asymptotics for $\sigma_{n}(x,t)$ never vanishes. Therefore, when $a_{2m+1}$ is large and $(x, t)$ in the neighborhood of $\left(\hat{x}_{0}, \hat{t}_{0}\right)$, we get from (\ref{sigmanxt5}) that
\[
u_N(x,t) = \frac{\sigma_{1}}{\sigma_{0}}e^{\textrm{i}t} =e^{\textrm{i}t}
\left(1- \frac{4[1+2\textrm{i}(t-\hat{t}_{0})]}{1+4(x-\hat{x}_{0})^2+4(t-\hat{t}_{0})^2}\right) + O\left(a_{2m+1}^{-1/(2m+1)}\right),
\]
which is a Peregrine wave $\hat{u}_1(x-\hat{x}_{0}, t-\hat{t}_{0}) \hspace{0.05cm} e^{\textrm{i}t}$, and the error of this Peregrine prediction is $O\left(a_{2m+1}^{-1/(2m+1)}\right)$. Theorem~3 is then proved.

\textbf{Proof of Theorem 4.} \hspace{0.05cm}
To analyze the large-$a_{2m+1}$ behavior of the rogue wave $u_N(x,t)$ in the neighborhood of the origin, where $\sqrt{x^2+t^2}=O(1)$, we first rewrite the $\sigma_n$ determinant (\ref{sigma_n}) into a $3N\times 3N$ determinant \cite{OhtaJY2012}
\[ \label{3Nby3Ndet2}
\sigma_{n}=\left|\begin{array}{cc}
\textbf{O}_{N\times N} & \Phi_{N\times 2N} \\
-\Psi_{2N\times N} & \textbf{I}_{2N\times 2N} \end{array}\right|,
\]
where $\Phi_{i,j}=2^{-(j-1)} S_{2i-j}\left[\textbf{\emph{x}}^{+}(n) + (j-1) \textbf{\emph{s}}\right]$, and
$\Psi_{i,j}=2^{-(i-1)} S_{2j-i}\left[\textbf{\emph{x}}^{-}(n) + (i-1) \textbf{\emph{s}}\right]$.
Defining $\textbf{\emph{y}}^{\pm}$ to be the vector $\textbf{\emph{x}}^{\pm}$ without the $a_{2m+1}$ term, i.e., let
\[
\textbf{\emph{x}}^{+}=\textbf{\emph{y}}^{+}+(0, \cdots, 0, a_{2m+1}, 0, \cdots), \quad
\textbf{\emph{x}}^{-}=\textbf{\emph{y}}^{-}+(0, \cdots, 0, a_{2m+1}^*, 0, \cdots),
\]
we find that the Schur polynomials of $\textbf{\emph{x}}^{\pm}$ are related to those of $\textbf{\emph{y}}^{\pm}$ as
\begin{equation} \label{Sjrelation}
S_{j}(\textbf{\emph{x}}^{+}+\nu\textbf{\emph{s}}) = \sum_{i=0}^{\left[\frac{j}{2m+1}\right]} \frac{a_{2m+1}^i}{i!} S_{j-(2m+1)i}(\textbf{\emph{y}}^{+}+\nu\textbf{\emph{s}}), \quad
S_{j}(\textbf{\emph{x}}^{-}+\nu\textbf{\emph{s}}) = \sum_{i=0}^{\left[\frac{j}{2m+1}\right]} \frac{(a_{2m+1}^*)^i}{i!} S_{j-(2m+1)i}(\textbf{\emph{y}}^{-}+\nu\textbf{\emph{s}}),
\end{equation}
where $[a]$ represents the largest integer less than or equal to $a$. Using this relation, we express matrix elements of $\Phi$ and $\Psi$ in Eq. (\ref{3Nby3Ndet2}) through Schur polynomials $S_{k}(\textbf{\emph{y}}^{\pm}+\nu\textbf{\emph{s}})$ and powers of $a_{2m+1}$ and $a_{2m+1}^*$.

We need to determine the highest power term of $a_{2m+1}$ in the determinant (\ref{3Nby3Ndet2}). For that purpose, it may be tempting to retain only the highest power term of $a_{2m+1}$ and $a_{2m+1}^*$ in each element of this determinant. That does not work though because it would result in multiple rows (and columns) which are proportional to each other, making the reduced determinant zero. The correct way is to first judiciously remove certain leading power terms of $a_{2m+1}$ and $a_{2m+1}^*$ from elements of the determinant through row and column manipulations, so that the remaining determinant, after retaining only the highest power term of $a_{2m+1}$ and $a_{2m+1}^*$ in each element, would be nonzero. These row and column manipulations are described below.

Suppose $N\equiv N_{0} \hspace{0.1cm} \mbox{mod} \hspace{0.1cm} (2m+1)$, i.e., $N=k(2m+1)+N_0$ for some positive integer $k$, with $0\le N_0\le m$. We perform the following series of row operations to the matrix $\Phi$ so that certain high-power terms of $a_{2m+1}$ in its lower rows are eliminated. In the first round, we use the 1st to $m$-th rows of $\Phi$ to eliminate the highest-power term $a_{2m+1}^{2\nu}$ from the $[\nu (2m+1) +1]$-th up to the $[\nu (2m+1) +m]$-th rows for each $1\le \nu \le k$, so that the remaining terms in those rows have the highest power $a_{2m+1}^{2\nu-1}$. We also use the $(m+1)$-th to $(2m+1)$-th rows of $\Phi$ to eliminate the highest-power term $a_{2m+1}^{2\nu+1}$ from the $[\nu (2m+1) +m+1]$-th to the $[\nu (2m+1) +2m+1]$-th rows for each $1\le \nu \le k-1$, with the remaining terms in those rows having the highest power $a_{2m+1}^{2\nu}$.
In each step, the highest power terms $a_{2m+1}^{2\nu}$ or $a_{2m+1}^{2\nu+1}$ of each row are eliminated simultaneously, because the coefficient vector of those highest power terms in each row below the $(2m+1)$-th is proportional to the coefficient vector of the highest power terms in the corresponding upper row between the 1st and $(2m+1)$-th due to the relation (\ref{Sjrelation}).

In the second round, we use the $(2m+1+1)$-th to $(2m+1+m)$-th rows of the remaining matrix $\Phi$ to eliminate the highest-power term $a_{2m+1}^{2\nu+1}$ from the $[(\nu+1)(2m+1)+1]$-th up to the $[(\nu+1)(2m+1)+m]$-th rows for each $1\le \nu \le k-1$, so that the remaining terms in those rows have the highest power $a_{2m+1}^{2\nu}$. We also use the $(2m+1+m+1)$-th to $(2m+1+2m+1)$-th rows of $\Phi$ to eliminate the highest-power term $a_{2m+1}^{2\nu+2}$ from the $[(\nu+1)(2m+1) +m+1]$-th up to the $[(\nu+1)(2m+1) +2m+1]$-th rows for each $1\le \nu \le k-2$, with the remaining terms in those rows having the highest power $a_{2m+1}^{2\nu+1}$. This process is repeated $k$ rounds.

At the end of this process, the $i$-th row of the remaining matrix $\Phi$ has the highest power $a_{2m+1}^{[(i+m)/(2m+1)]}$. Then, we keep only the highest power terms of $a_{2m+1}$ in each row. Similar column operations are also performed on the matrix $\Psi$. With these manipulations, we find that $\sigma_n$ is asymptotically reduced to
\[ \label{3Nby3Ndetm}
\sigma_{n}=\beta \hspace{0.06cm} |a_{2m+1}|^{k^2(2m+1)+k(2N_0+1)}\left|\begin{array}{cc}
\textbf{O}_{N\times N} & \widetilde{\Phi}_{N\times 2N} \\
-\widetilde{\Psi}_{2N\times N} &  \textbf{I}_{2N \times 2N }
\end{array}\right| \left[1+O\left(a_{2m+1}^{-1}\right)\right],
\]
where $\beta$ is a $(m,N)$-dependent nonzero constant, matrices $\widetilde{\Phi}_{N\times 2N}$ and $\widetilde{\Psi}_{2N\times N}$ have the structures
\[
\widetilde{\Phi}_{N\times 2N}=\left(\begin{array}{ccc} {\textbf{L}}_{(N-N_0)\times (N-N_0)} & \textbf{O}_{(N-N_0)\times 2N_0} & \textbf{O}_{(N-N_0)\times (N-N_0)} \\
{\textbf{M}}_{N_0\times (N-N_0)} & \widehat{\Phi}_{N_0\times 2N_0} & \textbf{O}_{N_0\times (N-N_0)} \end{array}\right),
\]
\[
\widetilde{\Psi}_{2N\times N}=\left(\begin{array}{cc} {\textbf{U}}_{(N-N_0)\times (N-N_0)} & \widehat{{\textbf{M}}}_{(N-N_0)\times N_0} \\
\textbf{O}_{2N_0\times (N-N_0)} & \widehat{\Psi}_{2N_0\times N_0} \\
\textbf{O}_{(N-N_0)\times (N-N_0)} & \textbf{O}_{(N-N_0) \times N_0} \end{array}\right),
\]
\[
{\textbf{L}}_{i,j}=S_{i-j}\left[\textbf{\emph{y}}^{+}+(j-1)\textbf{\emph{s}}\right], \quad
{\textbf{U}}_{i,j}=S_{j-i}\left[\textbf{\emph{y}}^{-}+(i-1)\textbf{\emph{s}}\right],
\]
\begin{eqnarray} \label{widehatPhiPsi}
\widehat{\Phi}_{i,j}=2^{-(j-1)} S_{2i-j}\left[\textbf{\emph{y}}^{+}(n) + (j-1+\nu_0) \textbf{\emph{s}}\right], \quad \widehat{\Psi}_{i,j}=2^{-(i-1)} S_{2j-i}\left[\textbf{\emph{y}}^{-}(n) + (i-1+\nu_0) \textbf{\emph{s}}\right],
\end{eqnarray}
$\nu_0= k(2m+1)$, and ${\textbf{M}}$, $\widehat{{\textbf{M}}}$ are matrices of elements $S_{j}(\textbf{\emph{y}}^{+}+\nu\textbf{\emph{s}})$ and $S_{j}(\textbf{\emph{y}}^{-}+\nu\textbf{\emph{s}})$ respectively. Since ${\textbf{L}}$ and ${\textbf{U}}$ are respectively lower triangular and upper triangular matrices with unit elements on the diagonal in view that $S_0=1$ and $S_j=0$ for $j<0$, $\sigma_n$ in Eq.~(\ref{3Nby3Ndetm}) then is
\[ \label{3Nby3Ndet3}
\sigma_{n}=\beta \hspace{0.06cm} |a_{2m+1}|^{k^2(2m+1)+k(2N_0+1)}
\left|\begin{array}{cc}
\textbf{O}_{N_0\times N_0} & \widehat{\Phi}_{N_0\times 2N_0} \\
-\widehat{\Psi}_{2N_0\times N_0} & \textbf{I}_{2N_0\times 2N_0} \end{array}\right| \left[1+O\left(a_{2m+1}^{-1}\right)\right].
\]
Finally, we notice that $S_j\left[\textbf{\emph{y}}^{\pm} + (\nu+\nu_0) \textbf{\emph{s}}\right]$ is related to $S_j\left(\textbf{\emph{y}}^{\pm} + \nu \textbf{\emph{s}}\right)$ through
\begin{equation}
S_{j}\left[\textbf{\emph{y}}^{\pm} + (\nu+\nu_{0}) \textbf{\emph{s}}\right]=\sum_{i=0}^{\left[j/2\right]}S_{2i}(\nu_{0}\textbf{\emph{s}}) S_{j-2i}(\textbf{\emph{y}}^{\pm} +\nu\textbf{\emph{s}}).
\end{equation}
Using this relation, the determinant in (\ref{3Nby3Ndet3}) can be reduced to one where $\nu_0$ is set to zero in the above $\widehat{\Phi}$ and $\widehat{\Psi}$ matrices given in Eq.~(\ref{widehatPhiPsi}). Such a determinant for $\sigma_n$ gives a $N_0$-th order rogue wave, whose internal parameters $(a_3, a_5, \cdots, a_{2N_0-1})$ are the first $N_0-1$ values in the original parameter set $(a_3, a_5, \cdots, a_{2N-1})$. Thus, in the neighborhood of the origin ,
\[ \label{uNuN0}
u_N(x,t; a_3, a_5, \cdots, a_{2N-1}) = \frac{\sigma_{1}}{\sigma_{0}}e^{\textrm{i}t}= u_{N_{0}}(x,t; a_{3}, a_{5}, \cdots, a_{2N_{0}-1}) \left[1+O\left(a_{2m+1}^{-1}\right)\right], \quad |a_{2m+1}|\gg 1,
\]
which means that the original $N$-th order rogue wave $u_N(x,t)$ is approximated by a lower $N_0$-th order rogue wave $u_{N_0}(x,t)$, with the approximation error $O\left(a_{2m+1}^{-1}\right)$.

If $N\equiv -N_{0}-1 \hspace{0.1cm} \mbox{mod} \hspace{0.1cm} (2m+1)$ with $0\le N_0\le m$, Eq.~(\ref{uNuN0}) can also be derived by similar analysis, and thus the same conclusion holds.

Lastly, we recall that $1\le m\le N-1$. In addition, $0\le N_0\le m$ in view of Theorem 2. Furthermore, when $m=N-1$, we find from Eq.~(\ref{N02}) of Theorem 2 that $N_0=N-2$. As a consequence, $0\le N_0\le N-2$. Theorem 4 is then proved.

\section{Rogue wave patterns when multiple internal parameters are large} \label{sec:gen}
The rogue patterns in Theorems~3 and 4 were derived under the assumption that only one of the internal parameters in the rogue wave solutions (\ref{BilinearTrans2}) was large, and the other parameters were $O(1)$. It turns out that those results can be generalized to more general parameter conditions. We discuss these generalizations in this section.

Regarding the generalization of Theorem~3, we can show that if $a_{2m+1}$ is large, and the other parameters $a_3$, $\dots$, $a_{2m-1}$, $a_{2m+3}$, $\dots$, $a_{2N-1}$ are also large but satisfy the conditions
\[ \label{acond}
a_{2j+1}=o\left( a_{2m+1}^{\frac{2j}{2m+1}} \right),\quad j\neq m,
\]
then, far away from the origin, with $\sqrt{x^2+t^2}=O\left(|a_{2m+1}|^{1/(2m+1)}\right)$, the rogue wave $u_N(x,t)$ still separates into $N_p$ Peregrine waves, whose positions $(\hat{x}_{0}, \hat{t}_{0})$ are given by Eq.~(\ref{x0t0}). Expressed mathematically, when $\left[(x-\hat{x}_{0})^2+(t-\hat{t}_{0})^2\right]^{1/2}=O(1)$, we have
\[ \label{Theorem3asymb}
u_{N}(x,t; a_{3}, a_{5}, \cdots, a_{2N-1}) \longrightarrow \hat{u}_1(x-\hat{x}_{0},t-\hat{t}_{0})\hspace{0.05cm} e^{\textrm{i}t} \hspace{0.4cm} \mbox{as} \hspace{0.15cm} |a_{2m+1}| \to  \infty.
\]
The reason is that, when $\sqrt{x^2+t^2}=O\left(|a_{2m+1}|^{1/(2m+1)}\right)$, under the same notation $\lambda=a_{2m+1}^{-1/(2m+1)}$ as in Eq.~(\ref{deflambda}), the condition (\ref{acond}) means that $a_{2j+1}=o(\lambda^{-2j})$ for $j\ne m$. Thus, $x_{2j+1}^+\lambda^{2j+1}=o(\lambda)$ for $j\ne m$. Then, in view of Eq.~(\ref{Skdef}), we have
\begin{eqnarray} \label{Skasymf}
&& S_{k}(\textbf{\emph{x}}^{+}(n) +\nu \textbf{\emph{s}}) =
S_{k}\left(x_{1}^{+}, \nu s_{2}, x_{3}^{+}, \nu s_4, \cdots\right)=\lambda^{-k} S_{k}\left(x_{1}^{+}\lambda, \nu s_{2}\lambda^2, x_{3}^{+}\lambda^3, \nu s_{4}\lambda^4, \cdots\right) \nonumber \\
&&= \lambda^{-k} S_{k}\left(x_{1}^{+}\lambda, 0, \cdots, 0, 1, 0, \cdots\right) \left[1+o(\lambda)\right]
=S_{k}\left(x_{1}^{+}, 0, \cdots, 0, a_{2m+1}, 0, \cdots\right)\left[1+o(\lambda)\right].
\end{eqnarray}
This relation is the counterpart of Eq.~(\ref{Skasym2}) in the proof of Theorem~3. Due to this relation and a similar one on $S_{k}(\textbf{\emph{x}}^{-}(n) +\nu \textbf{\emph{s}})$, the calculations in the proof of Theorem~3 can still go through. The only difference is that the error of the present Peregrine approximation may be different. Indeed, the previous analysis, combined with the above equation (\ref{Skasymf}), indicates that the error of the current Peregrine approximation (\ref{Theorem3asymb}) is the largest order among
$O\left(a_{2j+1}/a_{2m+1}^{2j/(2m+1)}\right)$, where $1\le j\le N-1$ and $j\ne m$. So, if $a_{2j+1}=O\left( a_{2m+1}^{(2j-1)/(2m+1)} \right)$ or smaller for all $j\ne m$, then the error of the current Peregrine approximation (\ref{Theorem3asymb}) would remain the same as that given in Eq.~(\ref{Theorem3asym}) of Theorem~3, i.e., $O\left(|a_{2m+1}|^{-1/(2m+1)}\right)$. Otherwise, this error would be larger than $O\left(|a_{2m+1}|^{-1/(2m+1)}\right)$, which means that the error would decay to zero slower than the rate $|a_{2m+1}|^{-1/(2m+1)}$ when $a_{2m+1}$ gets large.

Regarding the generalization of Theorem~4, we can show that if $a_{2m+1}$ is large, and
\[ \label{acond2}
a_3, \cdots, a_{2m-1} =O(1), \quad a_{2m+3}, \cdots, a_{2N-1}=O(a_{2m+1}),
\]
then Theorem~4 remains valid. Specifically, the asymptotics (\ref{ucenter}), including its error estimates, still holds. The proof for this is an extension of the proof for Theorem~4, and will be presented in Appendix C.

\begin{figure}[htb]
\begin{center}
\includegraphics[scale=0.84, bb=200 0 385 170]{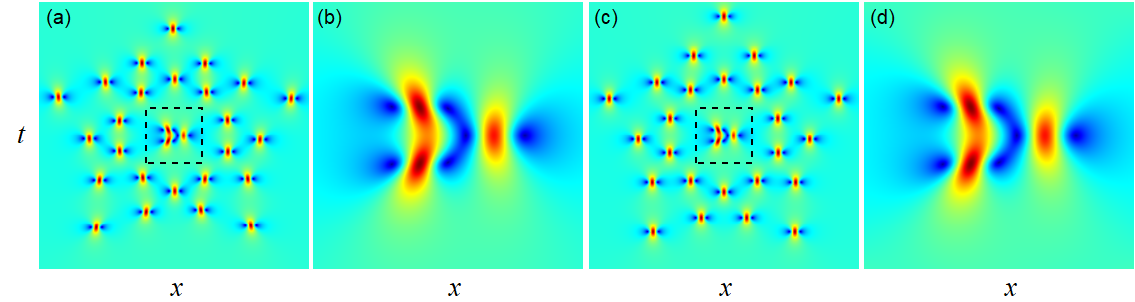}
\caption{A 7th order rogue wave $|u_7(x,t)|$ for generalized parameters (\ref{a3a13}) with $a_5=-200\textrm{i}$, and its comparison with the analytical prediction. (a) True solution, with $ -20.5\le x, t\le 20.5$; (b)
zoomed-in plot of the center region of the true solution marked by a dashed-line box in panel (a); (c) predicted solution with the same $(x, t)$ internals as in (a); (d) zoomed-in plot of the center region of the predicted solution.  }   \label{f:multiple}
\end{center}
\end{figure}

To demonstrate these generalized results on rogue patterns, we consider an example of a 7th order rogue wave $u_7(x,t)$ with parameter choices of
\[ \label{a3a13}
a_3=1, \quad a_5\ \mbox{is large}, \quad a_7=a_5, \quad a_9=2a_5, \quad a_{11}=3a_5, \quad a_{13}=4a_5.
\]
This set of parameters satisfy both conditions (\ref{acond}) and (\ref{acond2}). Thus, according to the above discussions, both Theorems~3 and 4 remain valid, including their error estimates, since $a_{2j+1}=O\left( a_{2m+1}^{(2j-1)/(2m+1)} \right)$ or smaller for all $j\ne m$ here. These theorems predict that far away from the origin, this $u_7(x,t)$ would split into 25 Peregrine waves, whose $(x,t)$ locations are given by Eq.~(\ref{x0t0}). Near the origin, this $u_7(x,t)$ would reduce to a 2nd-order rogue wave $u_2(x,t)$ with $a_3=1$. To verify these predictions, we choose $a_5=-200\textrm{i}$. The corresponding true rogue wave solution $|u_7(x,t)|$ is plotted in Fig.~\ref{f:multiple}(a), and its center region is amplified and shown in panel (b). Our asymptotic predictions (\ref{upNLS}) from Theorems~3 and 4 for the same $(x,t)$ intervals as in panels (a) and (b) are displayed in panels (c) and (d) respectively. One can clearly see that our predictions are almost indistinguishable from the true solutions.

\vspace{-0.5cm}
\section{Conclusions and discussions}
In this paper, we have analytically studied rogue wave patterns in the NLS equation. We have shown that when one of the internal parameters in the bilinear rogue wave solutions is large, these waves would exhibit clear geometric structures, which comprise Peregrine rogue waves organized in shapes such as triangle, pentagon, heptagon and nonagon, with a possible lower-order rogue wave at its center. These rogue patterns are analytically determined by the root structures of the Yablonskii-Vorob'ev polynomial hierarchy, and their orientations are controlled by the phase of the large parameter. We have also generalized these results and shown that, when multiple internal parameters in the rogue waves are large but satisfy certain constraints [such as (\ref{acond}) and (\ref{acond2})], then the same rogue patterns would persist. Comparison between true rogue patterns and our analytical predictions has shown excellent agreement. As a small application of our analytical results, the numerical observation in \cite{KAAN2011} on single-shell ring structures has been explained. Our results reveal the deep connection between NLS rogue wave patterns and the Yablonskii-Vorob'ev polynomial hierarchy, and make prediction of sophisticated patterns in higher-order NLS rogue waves possible.

It turns out that this connection between rogue wave patterns and the Yablonskii-Vorob'ev polynomial hierarchy is not restricted to the NLS equation. We have found that such connections persist in many other integrable equations, such as the derivative NLS equation, the Boussinesq equation, the Manakov equations and others. This general connection then gives rise to universal rogue wave patterns in integrable systems. This university result was briefly reported in \cite{YangYanguniv}. Its details are beyond the scope of this paper and will be pursued in future publications.

In this article, NLS rogue wave patterns are determined by the complex roots of the Yablonskii-Vorob'ev polynomial hierarchy, and these roots are the pole locations of rational solutions to the $\mbox{P}_{\mbox{\scriptsize II}}$ hierarchy (see Sec.~\ref{secPII} and \cite{Clarkson2003-II,Bertola2016}).
Interestingly, in very different contexts, somewhat similar results have also been reported. For instance, in the semiclassical NLS equation after wave breaking, a sequence of Peregrine waves appear, and their locations are determined by the poles of the tritronqu\'ee solution to the first Painlev\'e
($\mbox{P}_{\mbox{\scriptsize I}}$) equation \cite{Tovbis}. In the semiclassical sine-Gordon equation with initial conditions near the separatrix of a simple pendulum, superluminal (infinite velocity) kinks that appear in the solution are linked to the \emph{real} roots of the Yablonskii-Vorob'ev polynomials associated with rational solutions of the $\mbox{P}_{\mbox{\scriptsize II}}$ equation \cite{Miller_sine}. This connection of wave phenomena to rational solutions of the Painlev\'e equations may arise again in other wave systems in the future.
\section*{Acknowledgment}
This material is based upon work supported by the National Science Foundation under award number DMS-1910282, and the Air Force Office of Scientific Research under award number FA9550-18-1-0098.

\section*{Appendix A}
In this appendix, we briefly derive the bilinear rogue waves presented in Theorem 1. These new rogue wave expressions can be obtained by applying a new parameterization developed in Ref.~\cite{YangDNLS2019} to the bilinear derivation of rogue waves in Ref.~\cite{OhtaJY2012}. Specifically, instead of the previous choice (3.11) for the matrix element $m_{ij}^{(n)}$ in Ref.~\cite{OhtaJY2012}, which we denote as $\phi_{ij}^{(n)}$ in this paper, we now choose
\[ \label{mijdiff}
\phi_{ij}^{(n)}=\left. \frac{1}{i!}(p\partial_p)^i \frac{1}{j!}(q\partial_q)^j \phi^{(n)}\right|_{p=q=1},
\]
where
\[
\phi^{(n)}=\frac{(p+1)(q+1)}{2(p+q)}\left(-\frac{p}{q}\right)^n \exp\left(\xi+\eta+\sum_{k=1}^\infty a_k (\ln p)^k +\sum_{k=1}^\infty b_k (\ln q)^k\right),
\]
\[
\xi=px_1+p^2x_2, \quad \eta=qx_1-q^2x_2.
\]
and $a_k, b_k$ are arbitrary complex constants. Obviously, the function $\tau_n=\det_{1\le i,j\le N}\left(\phi_{2i-1,2j-1}^{(n)}\right)$ with the above choice of $\phi_{ij}^{(n)}$ also satisfies the bilinear equations (3.14) in \cite{OhtaJY2012}. Then, when we set $b_k=a_k^*$, $x_1=x$ and $x_2=\textrm{i}t/2$, this $\tau_n$ function would satisfy the bilinear equations (3.1) of the NLS equation in \cite{OhtaJY2012} [with $t$ switched to $-t/2$ since the NLS equation (\ref{NLS-2020}) in this paper differs from that in \cite{OhtaJY2012} by this $t$ rescaling]. Applying the same reduction technique of \cite{OhtaJY2012} to the above new $\tau_n$ solution, we can remove the differential operators in the expression (\ref{mijdiff}) of its matrix element $\phi_{ij}^{(n)}$ and reduce it to $\sigma_n=\det_{1\le i,j\le N}\left(\phi_{2i-1,2j-1}^{(n)}\right)$, where
\begin{equation}
\phi_{i,j}^{(n)}=\sum_{\nu=0}^{\min(i,j)} \frac{1}{4^{\nu}} \hspace{0.06cm} S_{i-\nu}(\hat{\textbf{\emph{x}}}^{+}(n) +\nu \textbf{\emph{s}})  \hspace{0.06cm} S_{j-\nu}(\hat{\textbf{\emph{x}}}^{-}(n) + \nu \textbf{\emph{s}}),
\end{equation}
vectors $\hat{\textbf{\emph{x}}}^{\pm}(n)=\left( x_{1}^{\pm}, x_{2}^{\pm},\cdots \right)$ are defined by
\begin{eqnarray}
x_{1}^{+}=x + \textrm{i} t + n+a_1, \quad x_{1}^{-}=x -\textrm{i} t - n+a_1^*,
\quad x_{k}^{+}= \frac{x+2^{k-1} (\textrm{i} t)}{k!} +a_{k},    \quad x_{k}^{-}=  \frac{x-2^{k-1} (\textrm{i} t)}{k!}+ a_{k}^*, \quad k\ge 2,
\end{eqnarray}
and $\textbf{\emph{s}}=(s_1, s_2, \cdots)$ are coefficients from the expansion (\ref{sexpand}). Through a shift of the $x$ and $t$ axes, we normalize $a_1=0$ without loss of generality. Finally, we split the vectors $\hat{\textbf{\emph{x}}}^{\pm}(n)$ into $\textbf{\emph{x}}^{\pm}(n)+\textbf{\emph{w}}^{\pm}$, where $\textbf{\emph{x}}^{\pm}(n)$ is as given in Eq.~(\ref{xpmdef}), and $\textbf{\emph{w}}^{\pm}=(0, x_{2}^{\pm}, 0, x_{4}^{\pm}, \cdots)$. Since $\hat{\textbf{\emph{x}}}^{\pm}(n) +\nu \textbf{\emph{s}}=\textbf{\emph{x}}^{\pm}(n) +\nu \textbf{\emph{s}} + \textbf{\emph{w}}^{\pm}$, it is easy to show from the definition of Schur polynomials (\ref{Schurdef}) that
\[
S_{k}(\hat{\textbf{\emph{x}}}^{\pm}(n) +\nu \textbf{\emph{s}})=\sum_{j=0}^{\left[ k/2 \right]} S_{j}(\hat{\textbf{\emph{w}}}^\pm) S_{k-2j}(\textbf{\emph{x}}^{\pm}(n)+\nu \textbf{\emph{s}}),
\]
where $\hat{\textbf{\emph{w}}}^\pm=(x_{2}^{\pm}, x_{4}^{\pm}, \cdots)$. Rewriting the $\sigma_n$ solution $\det_{1\le i,j\le N}\left(\phi_{2i-1,2j-1}^{(n)}\right)$ as a $3N\times 3N$ determinant (\ref{3Nby3Ndet2})
and utilizing the above relation, we can apply row and column manipulations to eliminate all terms involving $\hat{\textbf{\emph{w}}}^\pm$ in this $3N\times 3N$ determinant. The remaining $3N\times 3N$ determinant then becomes $\det_{1\le i,j\le N}\left(\phi_{2i-1,2j-1}^{(n)}\right)$, whose matrix element $\phi_{ij}^{(n)}$ is as given in Theorem~1.

\section*{Appendix B}
In this appendix, we prove Theorem 2. First, we derive the multiplicity of root zero in $Q_{N}^{[m]}(z)$. For this purpose, we define the Schur polynomial $S^{[m]}_k(z; a)$ as
\begin{equation} \label{Skza}
\sum_{k=0}^{\infty}S^{[m]}_k(z; a) \epsilon^k
=\exp\left[z\epsilon +a \hspace{0.04cm} \epsilon^{2m+1}\right],
\end{equation}
where $a$ is a constant. Through these Schur polynomials $S^{[m]}_k(z; a)$, we define polynomials
\begin{eqnarray} \label{PNza}
&& P^{[m]}_{N} (z; a) = c_{N} \left| \begin{array}{cccc}
         S^{[m]}_{1}(z; a) & S^{[m]}_{0}(z; a) & \cdots &  S^{[m]}_{2-N}(z; a) \\
         S^{[m]}_{3}(z; a) & S^{[m]}_{2}(z; a) & \cdots &  S^{[m]}_{4-N}(z; a) \\
        \vdots& \vdots & \vdots & \vdots \\
         S^{[m]}_{2N-1}(z; a) & S^{[m]}_{2N-2}(z; a) & \cdots &  S^{[m]}_{N}(z; a)
       \end{array}
 \right|,
\end{eqnarray}
where $S^{[m]}_k(z; a)\equiv 0$ when $k< 0$. It is easy to see that $S^{[m]}_k(z; a)$ is related to the polynomial $p_{k}^{[m]}(z)$ in Eq.~(\ref{pkmz}) as
\begin{equation}
S^{[m]}_k(z; a)=\hat{a}^{k/(2m+1)}p_{k}^{[m]}(\hat{z}),  \quad \hat{z}\equiv \hat{a}^{-1/(2m+1)}z,
\quad \hat{a}\equiv -\hspace{0.04cm} (2m+1) \hspace{0.04cm} 2^{-2m} \hspace{0.03cm} a.
\end{equation}
Thus, the polynomial $P^{[m]}_{N} (z; a)$ is related to the Yablonskii-Vorob'ev polynomial hierarchy $Q_{N}^{[m]}(z)$ in Eq.~(\ref{QNm}) as
\[
P^{[m]}_{N} (z; a)=\hat{a}^{\frac{N(N+1)}{2(2m+1)}} Q_{N}^{[m]}(\hat{z}).
\]
This equation tells us that every term in the polynomial $P^{[m]}_{N} (z; a)$ is a constant multiple of $z^ia^j$, where $i+(2m+1)j=N(N+1)/2$. Thus, to determine the multiplicity of the zero root $z=0$ in $Q_{N}^{[m]}(z)$, we need to determine the highest power term of $a$ in $P^{[m]}_{N} (z; a)$. To do so, we utilize the relation
\[ \label{Sjzarelation}
S^{[m]}_j(z; a)=\sum_{i=0}^{\left[\frac{j}{2m+1}\right]} \frac{a^i}{i! [j-i(2m+1)]!}z^{j-i(2m+1)},
\]
which can be derived by splitting the right side of Eq. (\ref{Skza}) into a product of two exponentials and then expanding both exponentials into Taylor series of $\epsilon$. Using this relation, we express the matrix elements in the determinant (\ref{PNza}) through powers of $z$ and $a$. Then, we need to obtain the highest power term of $a$ in the resulting determinant. This problem resembles the derivation of the highest power term of $a_{2m+1}$ in the $\sigma_n$ determinant (\ref{3Nby3Ndet2}) during the proof of Theorem~4, where a polynomial relation (\ref{Sjrelation}) similar to the above (\ref{Sjzarelation}) was used. In this resemblance, the matrix $P^{[m]}_{N} (z; a)$ here is the counterpart of the $\Phi_{N\times N}$ matrix in Eq. (\ref{3Nby3Ndet2}), $a$ here is the counterpart of $a_{2m+1}$ in Eq.~(\ref{Sjrelation}), and $z^j$ in the above equation (\ref{Sjzarelation})
is the counterpart of $S_j(\textbf{\emph{y}}^{+}+\nu\textbf{\emph{s}})$ in Eq.~(\ref{Sjrelation}). Performing the same row operations as described in Theorem~4 to remove certain leading $a$-power terms in the lower rows of the determinant (\ref{PNza}), we can show that the highest-power term of $a$ in $P^{[m]}_{N}(z; a)$ is
\begin{eqnarray}
\rho_{0}\ a^{\frac{N^2+N-N_{0}^2-N_0}{2(2m+1)}} \left|\begin{array}{cccc}
 z & 1  & \cdots & 0\\
\frac{z^3}{3!} &  \frac{z^2}{2!} & \cdots & 0 \\
\vdots & \vdots & \vdots & \vdots \\
\frac{z^{2N_{0}-1}}{(2N_{0}-1)!}   & \frac{z^{2N_{0}-2}}{(2N_{0}-2)!}   & \ldots & \frac{z^{N_{0}}}{N_{0} !}
                 \end{array}
\right|=\hat{\rho}_{0}\  a^{\frac{(N-N_{0})(N+N_{0}+1)}{2(2m+1)}} z^{N_{0}(N_{0}+1)/2},
\end{eqnarray}
where $N_0$ is as given in Theorem~2, and $\rho_0, \hat{\rho}_{0}$ are $(m, N)$-dependent nonzero constants. This shows that the lowest power of $z$ in all terms of $P^{[m]}_{N} (z; a)$ is $N_{0}(N_{0}+1)/2$. Then, setting $a=-2^{2m}/(2m+1)$ where $P^{[m]}_{N} (z; a)$ becomes $Q_{N}^{[m]}(z)$, the multiplicity of the zero root in $Q_{N}^{[m]}(z)$ is $N_{0}(N_{0}+1)/2$.

To prove the form (\ref{QNmform}) of the polynomial $Q_{N}^{[m]}(z)$, we notice from the definition (\ref{pkmz}) of the polynomial $p_{k}^{[m]}(z)$ that this polynomial admits the symmetry
\[
p_{k}^{[m]}(z)=\omega^{-k}p_{k}^{[m]}(\omega z),
\]
where $\omega$ is any one of the $(2m+1)$-th root of 1, i.e., $\omega^{2m+1}=1$. This symmetry of $p_{k}^{[m]}(z)$ leads to the symmetry of $Q_{N}^{[m]}(z)$ as
\[ \label{QNmzsym}
Q_{N}^{[m]}(z)=\omega^{-N(N+1)/2} Q_{N}^{[m]}(\omega z).
\]
Since the multiplicity of the zero root in $Q_{N}^{[m]}(z)$ is $N_{0}(N_{0}+1)/2$, let us write
\[
Q_{N}^{[m]}(z)=z^{N_0(N_0+1)/2}q_{N}^{[m]}(z),
\]
where $q_{N}^{[m]}(z)$ is a polynomial of $z$ with a nonzero constant term. The symmetry (\ref{QNmzsym}) of the polynomial $Q_{N}^{[m]}(z)$ induces a symmetry for $q_{N}^{[m]}(z)$ as
\[
q_{N}^{[m]}(z)=\omega^{(N_0^2+N_0-N^2-N)/2}q_{N}^{[m]}(\omega z).
\]
Since $N_0^2+N_0-N^2-N=(N_0-N)(N_0+N+1)$, and in view of the $N_0$ value given in Theorem~2, we see that $(N_0^2+N_0-N^2-N)/2$ is divisible by $2m+1$, which means $\omega^{(N_0^2+N_0-N^2-N)/2}=1$. Thus, the above equation reduces to
\[
q_{N}^{[m]}(z)=q_{N}^{[m]}(\omega z).
\]
This symmetry of $q_{N}^{[m]}(z)$ dictates that $q_{N}^{[m]}(z)$ can only be a polynomial of $z^{2m+1}$. Hence the form (\ref{QNmform}) of the polynomial $Q_{N}^{[m]}(z)$ is proved.

Lastly, we derive the degree of the polynomial $Q_{N}^{[m]}(z)$ from its definition (\ref{QNm}). Notice from Eq.~(\ref{pkmz}) that the highest-degree term of $p^{[m]}_{k}(z)$ is $z^k/k!$. Retaining only this highest-degree term of $p^{[m]}_{k}(z)$ in the determinant (\ref{QNm}) for $Q_{N}^{[m]}(z)$ and evaluating the simplified determinant by the same technique as that used in Ref.~\cite{OhtaJY2012}, we can readily show that the degree of the polynomial $Q_{N}^{[m]}(z)$ is $N(N+1)/2$. Thus, Theorem~2 is proved.

\section*{Appendix C}

In this appendix, we prove the generalization of Theorem 4 presented in Sec. \ref{sec:gen} when $a_{2m+1}$ is large and the other parameters satisfy the conditions (\ref{acond2}). In this parameter regime, let us denote
\[ \label{newacond2}
a_{2m+3}=\beta_{1} \hspace{0.04cm} a_{2m+1}, \hspace{0.2cm} a_{2m+5}=\beta_{2} \hspace{0.04cm} a_{2m+1}, \hspace{0.2cm} \cdots, \hspace{0.2cm} a_{2N-1}=\beta_{N-m-1} \hspace{0.04cm} a_{2m+1},
\]
where $\beta_{1}, \beta_{2}, \cdots, \beta_{N-m-1}$ are $O(1)$ constants. We first split the vectors  $\textbf{\emph{x}}^{\pm}$ as
\[
\textbf{\emph{x}}^{+}=\textbf{\emph{y}}^{+}+\textbf{\emph{a}}, \quad
\textbf{\emph{x}}^{-}=\textbf{\emph{y}}^{-}+\textbf{\emph{a}}^*,
\]
where $\textbf{\emph{a}}=(0, \cdots, 0, a_{2m+1}, 0,  a_{2m+3}, 0, \cdots  0, a_{2N-1})$. Then, the Schur polynomials of $\textbf{\emph{x}}^{\pm}$ are related to those of $\textbf{\emph{y}}^{\pm}$ as
\begin{equation} \label{Sjrelation2}
S_{j}(\textbf{\emph{x}}^{+}+\nu\textbf{\emph{s}}) = \sum_{i=0}^{j} S_{i}(\textbf{\emph{a}})  S_{j-i}(\textbf{\emph{y}}^{+}+\nu\textbf{\emph{s}}), \quad
S_{j}(\textbf{\emph{x}}^{-}+\nu\textbf{\emph{s}}) = \sum_{i=0}^{j} S_{i}^*(\textbf{\emph{a}}) S_{j-i}(\textbf{\emph{y}}^{-}+\nu\textbf{\emph{s}}).
\end{equation}
In view of the definition of $\textbf{\emph{a}}$ and the notations in (\ref{newacond2}), we readily see from the definition of Schur polynomials that $S_{i}(\textbf{\emph{a}})$ is a polynomial of $a_{2m+1}$ with coefficients dependent on $(\beta_1, \beta_2, \cdots)$, and its highest degree in $a_{2m+1}$ is $[i/(2m+1)]$, i.e., the largest integer less than or equal to $i/(2m+1)$. Then, after a little manipulation and rearranging terms in the above equations, we get
\[ \label{AppSj1}
S_{j}(\textbf{\emph{x}}^{+}+\nu\textbf{\emph{s}}) = \sum_{i=0}^{\left[\frac{j}{2m+1}\right]}  a_{2m+1}^i
\sum_{k=0}^{\left[\frac{j-(2m+1)i}{2}\right]} c_{i,k}^+(m, \mbox{\boldmath$\beta$})\  S_{j-(2m+1)i-2k}(\textbf{\emph{y}}^{+}+\nu\textbf{\emph{s}})
\]
and
\[ \label{AppSj2}
S_{j}(\textbf{\emph{x}}^{-}+\nu\textbf{\emph{s}}) = \sum_{i=0}^{\left[\frac{j}{2m+1}\right]}  (a_{2m+1}^*)^i
\sum_{k=0}^{\left[\frac{j-(2m+1)i}{2}\right]} c_{i,k}^{-}(m, \mbox{\boldmath$\beta$})\  S_{j-(2m+1)i-2k}(\textbf{\emph{y}}^{-}+\nu\textbf{\emph{s}}),
\]
where the coefficients $c_{i,k}^\pm$ are dependent on $m$ and the vector $\mbox{\boldmath$\beta$}=(\beta_1, \beta_2, \cdots)$, and $c_{i,0}^\pm(m,\mbox{\boldmath$\beta$})=1/i!$.

These two Schur polynomial relations (\ref{AppSj1})-(\ref{AppSj2}) are the counterparts of those in Eq. (\ref{Sjrelation}) during the proof of Theorem 4. Using these relations, we can perform similar row and column operations to the $3N\times 3N$ determinant in Eq.~(\ref{3Nby3Ndet2}) to eliminate certain high order powers of $a_{2m+1}$ and $a_{2m+1}^*$. The main difference is that, a little more such eliminations are required here, because to eliminate a certain power of $a_{2m+1}$ or $a_{2m+1}^*$ in $S_{j}(\textbf{\emph{x}}^{\pm}+\nu\textbf{\emph{s}})$, one needs to eliminate a linear combination of polynomials $S_{j-(2m+1)i-2k}(\textbf{\emph{y}}^{\pm}+\nu\textbf{\emph{s}})$ now in view of the above two Schur polynomial relations. However, these eliminations follow a clear and regular pattern, so that they can always be achieved. Another small difference is that here, the row and column operations will produce some additional lower power terms of $a_{2m+1}$ and $a_{2m+1}^*$. But those lower-power terms will eventually be discarded since we will retain only the highest $a_{2m+1}$ and $a_{2m+1}^*$ power terms in each row and column respectively. Therefore, these similar row and column operations will still asymptotically reduce $\sigma_n$ to the same determinant (\ref{3Nby3Ndetm}) as before, and hence the generalization of Theorem 4 stated in Sec. \ref{sec:gen} can be proved.


\section*{References}
\vspace{-0.5cm}
\begin{thebibliography}{10}
\bibitem{Ocean_rogue_review}
K. Dysthe, H.E. Krogstad and P. M\"uller,
``Oceanic rogue waves,"
Annu. Rev. Fluid Mech. 40, 287 (2008).

\bibitem{Pelinovsky_book}
C. Kharif, E. Pelinovsky and A. Slunyaev,
\emph{Rogue Waves in the Ocean} (Springer, Berlin, 2009).

\bibitem{Solli_Nature}
D.R. Solli, C. Ropers, P. Koonath and B. Jalali,
``Optical rogue waves",
Nature 450, 1054 (2007).

\bibitem{Wabnitz_book}
S. Wabnitz (Ed.), \emph{Nonlinear Guided Wave Optics: A testbed for extreme waves} (IOP Publishing, Bristol, UK, 2017).

\bibitem{Benney}
D.J. Benney and A.C. Newell,
``The propagation of nonlinear wave envelopes",
J. Math. Phys. 46, 133 (1967).

\bibitem{Peregrine}
D.H. Peregrine,
``Water waves, nonlinear Schr\"odinger equations and their solutions,"
J. Aust. Math. Soc. B 25, 16 (1983).

\bibitem{AAS2009}
N. Akhmediev, A. Ankiewicz and J.M. Soto-Crespo,
``Rogue waves and rational solutions of the nonlinear Schr\"odinger equation,"
Phys. Rev. E 80, 026601 (2009).

\bibitem{DGKM2010}
P. Dubard, P. Gaillard, C. Klein and V.B. Matveev,
``On multi-rogue wave solutions of the NLS equation and positon solutions of the KdV equation,"
Eur. Phys. J. Spec. Top. 185, 247 (2010).

\bibitem{KAAN2011}
D.J. Kedziora, A. Ankiewicz and N. Akhmediev,
``Circular rogue wave clusters,"
Phys. Rev. E  84, 056611 (2011).

\bibitem{GLML2012}
B.L. Guo, L.M. Ling and Q.P. Liu,
``Nonlinear Schr\"odinger equation: generalized Darboux transformation and rogue wave solutions,"
Phys. Rev. E 85, 026607 (2012).

\bibitem{OhtaJY2012}
Y. Ohta and J. Yang,
``General high-order rogue waves and their dynamics in the nonlinear Schr\"odinger equation,"
Proc. R. Soc. Lond. A 468, 1716 (2012).

\bibitem{Kaup_Newell}
D.J. Kaup and A.C. Newell,
``An exact solution for a derivative nonlinear Schr\"odinger equation,"
J. Math. Phys. 19, 798 (1978).

\bibitem{KN_Alfven1}
K. Mio, T. Ogino, K. Minami and S. Takeda,
``Modified nonlinear Schr\"{o}inger equation for Alfv\'en waves propagating along the magnetic field in cold plasmas,"
J. Phys. Soc. Jpn. 41, 265 (1976).

\bibitem{Wise2007}
J. Moses, B.A. Malomed and F.W. Wise,
``Self-steepening of ultrashort optical pulses without self-phase modulation",
Phys. Rev. A 76, 021802 (2007).

\bibitem{KN_rogue_2011}
S.W. Xu, J.S. He and L.H. Wang,
``The Darboux transformation of the derivative nonlinear Schr\"odinger equation,"
J. Phys. A 44, 305203 (2011).

\bibitem{KN_rogue_2013}
B.L. Guo, L.M. Ling and Q.P. Liu,
``High-order solutions and generalized Darboux transformations of derivative nonlinear Schr\"odinger equations,"
Stud. Appl. Math. 130, 317 (2013).

\bibitem{YangDNLS2019}
B. Yang, J. Chen and J. Yang,
``Rogue waves in the generalized derivative nonlinear Schr\"{o}dinger equations", J. Nonl. Sci. 30, 3027-3056 (2020).

\bibitem{Menyuk}
P.K.A. Wai and C.R. Menyuk,
``Polarization mode dispersion, decorrelation, and diffusion in optical fibers with randomly varying birefringence,"
J. Lightwave Technol. 14, 148 (1996).

\bibitem{BDCW2012}
F. Baronio, A. Degasperis, M. Conforti and S. Wabnitz,
``Solutions of the vector nonlinear Schr\"odinger equations: evidence for deterministic rogue waves",
Phys. Rev. Lett. 109, 044102 (2012).

\bibitem{ManakovDark}
F. Baronio, M. Conforti, A. Degasperis, S. Lombardo, M. Onorato and S. Wabnitz,
``Vector rogue waves and baseband modulation instability in the defocusing regime",
Phys. Rev. Lett. 113, 034101 (2014).

\bibitem{LingGuoZhaoCNLS2014}
L. Ling, B. Guo and L. Zhao, ``High-order rogue waves in vector nonlinear Schr\"{o}dinger equations", Phys Rev E, 89, 041201(R) (2014).

\bibitem{Chen_Shihua2015}
S. Chen and D. Mihalache, ``Vector rogue waves in the Manakov system: diversity and compossibility",
J. Phys. A 48, 215202 (2015).

\bibitem{ZhaoGuoLingCNLS2016}
L. Zhao, B. Guo and L. Ling, ``High-order rogue wave solutions for the coupled
nonlinear Schr\"odinger equations-II", J. Math. Phys. 57, 043508 (2016).

\bibitem{Ablowitz_book}
M.J. Ablowitz and H. Segur, \emph{Solitons and the Inverse Scattering Transform} (SIAM, Philadelphia, 1981).

\bibitem{BaroDegas2013}
F. Baronio, M. Conforti, A. Degasperis and S. Lombardo,
``Rogue waves emerging from the resonant interaction of three waves."
Phys. Rev. Lett. 111, 114101 (2013).

\bibitem{DegasLomba2013}
A. Degasperis and S. Lombardo,
``Rational solitons of wave resonant-interaction models",
Physical Review E, 88, 052914 (2013).

\bibitem{ChenSCrespo2015}
S. Chen, J.M. Soto-Crespo and P. Grelu,
``Watch-hand-like optical rogue waves in three-wave interactions",
Optics Express 23, 349-359 (2015).

\bibitem{WangXChenY2015}
X. Wang, J. Cao and Y. Chen,
``Higher-order rogue wave solutions of the three-wave resonant interaction equation via the generalized Darboux transformation",
Physica Scripta 90, 105201 (2015).

\bibitem{ZhangYanWen2018}
G. Zhang, Z. Yan and X.Y. Wen,
``Three-wave resonant interactions: Multi-dark-dark-dark solitons, breathers, rogue waves, and their interactions and dynamics",
Physica D 366, 27-42 (2018).

\bibitem{Tank1}
A. Chabchoub, N. Hoffmann and N. Akhmediev,
``Rogue wave observation in a water wave tank,"
Phys. Rev. Lett. 106, 204502 (2011).

\bibitem{Tank2}
A. Chabchoub, N. Hoffmann, M. Onorato, A. Slunyaev, A. Sergeeva, E. Pelinovsky and N. Akhmediev,
``Observation of a hierarchy of up to fifth-order rogue waves in a water tank,"
Phys. Rev. E 86, 056601 (2012).

\bibitem{Fiber1}
B. Kibler, J. Fatome, C. Finot, G. Millot, F. Dias, G. Genty, N. Akhmediev and J.M. Dudley,
``The Peregrine soliton in nonlinear fibre optics,"
Nat. Phys. 6, 790 (2010).

\bibitem{Fiber2}
B. Frisquet, B. Kibler, P. Morin, F. Baronio, M. Conforti, G. Millot and S. Wabnitz,
``Optical dark rogue wave,"
Sci. Rep. 6, 20785 (2016).

\bibitem{Fiber3}
F. Baronio, B. Frisquet, S. Chen, G. Millot, S. Wabnitz and B. Kibler,
``Observation of a group of dark rogue waves in a telecommunication optical fiber",
Phys. Rev. A 97, 013852 (2018).

\bibitem{HeFokas}
J.S. He, H.R. Zhang, L.H. Wang, K. Porsezian and A.S. Fokas,
``Generating mechanism for higher-order rogue waves",
Phys. Rev. E 87, 052914 (2013).

\bibitem{KAAN2013}
D.J. Kedziora, A. Ankiewicz and N. Akhmediev,
``Classifying the hierarchy of nonlinear-Schr\"{o}dinger-equation rogue-wave solutions."
Phys. Rev. E 88, 013207 (2013).

\bibitem{Ablowitzbook}
M.J. Ablowitz and H. Segur, \emph{Solitons and the Inverse Scattering Transform} (SIAM, Philadelphia, 1981).

\bibitem{Kivsharbook}
Y.S. Kivshar and G.P. Agrawal, \emph{Optical Solitons: From Fibers to Photonic Crystals}
(Academic Press, San Diego, 2003).

\bibitem{Yablonskii1959}
A.I. Yablonskii, Vesti Akad. Navuk. BSSR Ser. Fiz. Tkh. Nauk. 3, 30 (1959) (in Russian).

\bibitem{Vorobev1965}
A.P. Vorob'ev,
``On rational solutions of the second Painlev?equation",
Diff. Eqns. 1, 58 (1965).

\bibitem{Kajiwara-Ohta1996}
K. Kajiwara and Y. Ohta,
``Determinant structure of the rational solutions for the Painlev\'{e} II equation",
J. Math. Phys. 37, 4693 (1996).

\bibitem{Clarkson2003-II}
P.A. Clarkson and E.L. Mansfield,
``The second Painlev\'{e} equation, its hierarchy and associated special polynomials",
Nonlinearity 16, R1 (2003).

\bibitem{Bertola2016}
F. Balogh, M. Bertola and T. Bothner,
``Hankel determinant approach to generalized Vorob'ev-Yablonski polynomials and their roots",
Constr. Approx. 44, 417 (2016).

\bibitem{Fukutani}
S. Fukutani, K. Okamoto and H. Umemura,
``Special polynomials and the Hirota bilinear relations of the second and the fourth Painlev\'e equations",
Nagoya Math. J. 159, 179-200 (2000).

\bibitem{Taneda}
M. Taneda,
``Remarks on the Yablonskii-Vorob'ev polynomials",
Nagoya Math. J. 159, 87-111 (2000).

\bibitem{Miller2014}
R.J. Buckingham and P.D. Miller, ``Large-degree asymptotics of rational
Painlev\'e-II functions: noncritical behaviour", Nonlinearity 27, 2489 (2014).

\bibitem{YangYanguniv}
B. Yang and J. Yang, ``Universal patterns of rogue waves", arXiv:2009.06060 [nlin.PS] (2020).

\bibitem{Tovbis}
M. Bertola and A. Tobvis, ``Universality for the focusing nonlinear Schr\"odinger
equation at the gradient catastrophe point: rational breathers and poles of the
tritronqu\'ee solution to Painlev\'e I", Comm. Pure Appl. Math. 66, 678 (2013).

\bibitem{Miller_sine}
R. Buckingham and P.D. Miller, ``The sine-Gordon equation in the semiclassical limit: critical behavior near a separatrix," Journal d'Analyse Math\'{e}matique 118, 397-492 (2012).

\end{thebibliography}
\end{document}